\documentclass[aps,pra,showpacs,twoside,10pt,floatfix,nofootinbib,longbibliography,twocolumn]{revtex4-2}
\usepackage[colorlinks=true, citecolor=red, urlcolor=blue ]{hyperref}
\usepackage{epsfig,newlfont,amssymb,amsfonts,amsmath,bm,subfigure,palatino,mathtools,amsthm,braket,soul,enumitem,color,times,comment,geometry,float,array,makecell}
\usepackage[table]{xcolor}
\usepackage[normalem]{ulem}
\usepackage{silence}
\begin{document}

\title{
Developing universal logical state-purification strategy for quantum error correcting codes}
\author{Chandrima B. Pushpan$^1$, Tanoy Kanti Konar$^2$, Aditi Sen(De)$^2$, Amit Kumar Pal$^1$}
\affiliation{$^1$Department of Physics, Indian Institute of Technology Palakkad, Palakkad 678 623, India \\
$^2$Harish-Chandra Research Institute, A CI of Homi Bhabha National Institute, Chhatnag Road, Jhunsi, Prayagraj 211 019, India}

\begin{abstract}

We develop a measurement-based protocol for simultaneously purifying arbitrary logical states in multiple quantum error correcting codes with unit fidelity and finite probability, starting from arbitrary thermal states of each code. The protocol entails a time evolution caused by an engineered Hamiltonian, which results in transitions between the logical and error subspaces of the quantum error correcting code mediated by the auxiliary qubit, followed by a projective measurement in an optimum basis on the auxiliary qubit and an appropriate post-selection of the measurement outcomes. We  illustrate the results with the three-qubit repetition code and the logical qubit used in quantum state transfer protocol. We further demonstrate that when the measurement base is not optimal, it is possible to achieve both classical fidelity, and fidelity as high as $90\%$ through several iterations of the purifying procedure, thereby establishing its robustness against variations in the measurement basis. By repeating the purification rounds, we show that purifying the cardinal states of the logical Bloch sphere corresponding to logical qubits in quantum state transfer is feasible utilizing paradigmatic quantum spin models as the generator of the time evolution. 

\end{abstract}

\maketitle

\section{Introduction}
\label{sec:intro}

The boundary between the twentieth and the twenty-first century has been marked with the emergence of quantum technologies~\cite{Osada2022} promising to revolutionize the fields of communication and computation with the use of quantum mechanical principles~\cite{nielsen2010,*Hayashi2012,*Vedral2013,*Wilde_2013,Mermin2012,*Scherer2019}. For instance, this has led to studies on establishing secure quantum communication between distant parties \cite{Ekert91,*Bennett1992Crypto,*bennett1992,*bennett1993,quantum_crypto_gisin_rmp_2002}, and  solving complex physical problems using a quantum computer faster than what is possible with its classical counterpart \cite{Shor1997,Harrow2017,*Dalzell2020}. 
Experimental implementations of the theoretical results on quantum communication~\cite{Ekert91,*Bennett1992Crypto,*bennett1992,*bennett1993} and computation~\cite{Shor1997,raussendorf2001,*raussendorf2003} has been possible using photons~\cite{Mattle1996,*Bouwmeester1997,*Jennewein2000,*Walther_2005,*Lanyon2007,*Lu_2007,*Lu_2007a,Romero2024},  trapped ions~\cite{haffner2008,Lanyon_2013}, superconducting qubits~\cite{Arriagada_2018,Krantz2019,*Kjaergaard2020,*Bravyi2022}, and nuclear magnetic resonance (NMR) molecules~\cite{Vandersypen2001} (c.f.~\cite{Ladd2010Mar}) due to the realization of high-quality physical qubits -- the basic unit of quantum information processing -- with long coherence time in these substrates.

Processing quantum information in a falut-tolerant fashion~\cite{Shor1996a,*preskill1998,*gottesman2010} requires (a) encoding the information in the low-energy subspace of a quantum system via specitic interactions between multiple physical qubits, spanned by two chosen energy levels, termed as the \emph{logical subspace}~\cite{Shor1996a,*preskill1998,*gottesman2010,Terhal2015,*Roffe2019}, (b) ensuring the population to be confined within the logical subspace, such that the system works as a \emph{logical qubit}, and (c) protecting the logical subspace via error mitigation~\cite{Cai2023} and correction~\cite{Shor1996a,*preskill1998,gottesman2010,Terhal2015,*Roffe2019}. Theoretical construction of a logical qubit via assembling and controlling a number of physical qubits is extensively used in quantum error correcting (QEC) codes~~\cite{Shor1996a,*preskill1998,gottesman2010,Terhal2015,*Roffe2019}, including the Calderbank-Shor-Steane code~\cite{Calderbank1996,*Steane1996}, topological quantum codes~\cite{fujii2015,*prakash_pra_2015} such as the Kitaev code~\cite{kitaev2001,*kitaev2003,*kitaev2006} and the color code~\cite{bombin2006,*bombin2007}. The importance of such constructions is specifically highlighted in the ongoing  noisy intermediate-scale quantum (NISQ) era~\cite{Preskill_quantum_2018_nisq,*nash2020}, where QEC codes (QECCs) are being realized using physical platforms such as trapped ions~\cite{haffner2008} and superconducting qubits~\cite{Krantz2019,*Kjaergaard2020,*Bravyi2022} for proof of principle experiments with low-distance QECCs~\cite{Chiaverini2004,*Schindler2011,*Nigg2014,*Linke2017,*Pal_quantum_2022_relaxation,Reed2012,*Kelly2015,*Andersen2020,*Zhao2022,*Krinner2022May} as well as for implementing the envisioned fault-tolerant scalable quantum computers~\cite{Acharya2024,gao2024}. Idea similar to logical qubit has also been exploited in extending the quantum state transfer protocol~\cite{Bose03} to arbitrary qubit pairs on two-dimensional square lattice of arbitrary size~\cite{Pushpan2024Jul}.

A fundamental step towards fault-tolerant quantum computation is the initialization~\cite{Fowler2012,*Gidney2024} of logical qubit(s) in a \emph{fiducial state} -- typically a \emph{cardinal state} of the logical Bloch sphere --  appropriate for performing the QEC protocol. However, a logical qubit prepared in a desired state is often susceptible to different noise sources, including the thermal noise, resulting in \emph{leakage} from the logical to the error subspace (c.f.~\cite{Aliferis2007,*Miao2023}), which leads to the degradation of the quality of the quantum processor~\cite{Manabe2023,Acharya2024}.  Therefore, a post-initialization purification of the logical state is necessary, and yet a difficult feat (c.f. magic state distillation~\cite{Bravyi2005,*Li_2015,*Campbell2017,*Gidney2019,*Rodriguez2024}). 

\begin{figure}
    \centering
    \includegraphics[width=0.9\linewidth]{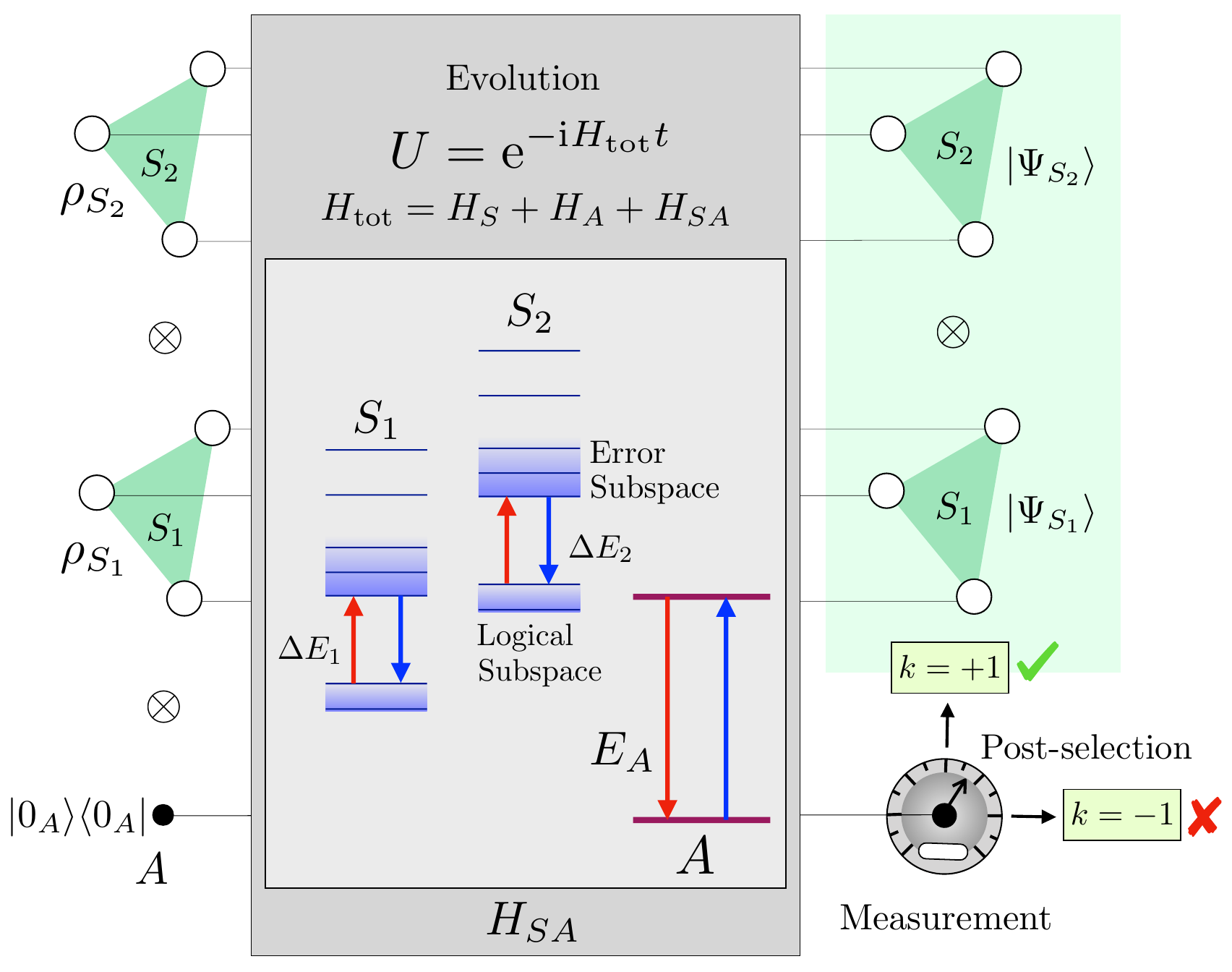}
    \caption{Schematic representation of the purification protocol for two QECCs $S_1$ and $S_2$ ($L=2$). Starting from arbitrary thermal states $\rho_{S_1}$ and $\rho_{S_2}$ on respectively $S_1$ and $S_2$, and the state $\ket{0_A}\bra{0_A}$ on the auxiliary qubit $A$, a time evolution $U=\exp-\text{i}H_{\text{tot}}t$, up to a time $t$, followed by a projective measurement on $A$ and an appropriate post-selection of the measurement outcome $k$ leads to perfect purification of arbitrary logical states $\ket{\Psi_{S_1}}$ and $\ket{\Psi_{S_2}}$ respectively on $S_1$ and $S_2$. The time-evolution is generated by turning on an interaction Hamiltonian $H_{SA}$ of the form (\ref{eq:interaction_degenerate_multiple_LQ}), which, for $E_A=\Delta E_1+\Delta E_2$, results in an energy-conserving transition between the logical and the error subspace.}
    \label{fig:schematics}
\end{figure}

In this work, we propose a measurement-based scheme to \emph{perfectly purify} an array of \emph{target} logical qubits with finite probability, starting from an arbitrary thermal state of each of the logical qubit with the aid of a fully accessible auxiliary qubit initialized at its ground state. We achieve this goal by \emph{designing} a specific interaction Hamiltonian that initiates a transition between the qubit eigenstates and the logical and the error subspaces of each logical qubit (see Fig.~\ref{fig:schematics}), resulting in a time evolution of the QECCs and the auxiliary qubit. We identify a measurement strategy carried out on the auxiliary qubit which, upon post-selection,  leads to the creation of the desired logical state(s) with unit fidelity and finite probability,  irrespective of the specific QECCs corresponding to the logical qubits.  

We further demonstrate that allowing multiple rounds of this purification protocol results in a boost in the purified logical state-fidelity corresponding to measurements for which the fidelity obtained after a single round is sub-optimal. The methodology applies to cases outside the domain of QEC also, eg. the logical qubits prepared for the purpose of quantum state transfer~\cite{Pushpan2024Jul} using paradigmatic quantum spin models. Our protocol also serves as a ground state preparation protocol for arbitrary Hamiltonians, and enjoys the efficiency of perfect purification with a single round of measurement, as opposed to the recently emerging measurement-based quantum state preparation protocols~\cite{jing_measure_ghz,Nakazato_2003,*Nakazato_2004,gefen_prr_2020,*gefen_prx_2024} for which multiple measurements are mandatory.

The rest of the paper is organized as follows. In Sec.~\ref{sec:designed_interaction}, we present the measurement-based protocol for perfect purification of the logical states with a finite probability. We also discuss the implications of different measurements on the auxiliary qubit on the performance of the protocol, and apply the protocol for specific QECCs and logical qubit implemented in isotropic Heisenberg model for demonstration. In Sec.~\ref{sec:repeated_measurements},  repeated purification rounds is employed  to achieve the goal. The concluding remarks  are presented in Sec.~\ref{sec:conclusion}.

\section{Measurement-based purification of logical states}
\label{sec:designed_interaction}

Consider a $D$-level quantum system $S$ representing a QECC, described by a stabilizer Hamiltonian $H_S$, satisfying
\begin{eqnarray}
H_S \ket{\mu_S}&=&E_{\mu}\ket{\mu_S}, \; \mu=0, 1, \cdots, D-1,
\label{eq:general_system_hamiltonian}
\end{eqnarray}
with \(\ket{\mu_S}\) being the eigenstate of $H_S$  corresponding to the eigenvalue \(E_{\mu}\). The Hamiltonian $H_S$ has a doubly-degenerate ground state that operates as the \emph{logical subspace} (LS), with ground state energy $E_0=E_1=0$, and an energy gap $\Delta E=E_2$, separating the LS spanned by  $\{\ket{0_S},\ket{1_S}\}$ from the higher energy \emph{error subspace} (ES),  starting from the first excited state $\ket{2_S}$ with a  $d$-fold degeneracy, such that $E_2=E_3=\cdots,E_{d+1}$, $d\geq 1$.  An  arbitrary logical state is given by $\ket{\Psi_S}=\cos{\frac{\theta}{2}}\ket{0_S}+\text{e}^{\text{i}\phi}\sin{\frac{\theta}{2}}\ket{1_S}$,  with $0\leq \theta\leq \pi$, and $0\leq\phi\leq 2\pi$, while the logical operators are defined as $X = \ket{0_S}\bra{1_S}+\ket{1_S}\bra{0_S}$, $Y = -\text{i}\left[\ket{1_S}\bra{0_S}-\ket{0_S}\bra{1_S}\right]$, and $Z =\ket{1_S}\bra{1_S}-\ket{0_S}\bra{0_S}$, where we have chosen $\{\ket{0_S}, \ket{1_S}\}$ as the logical $Z$ basis, and $\ket{0_S}$ as the ground state of the logical $Z$ operator.

Let us now consider $L$
QECCs, labeled by $S_1,S_2,\cdots,S_i$, each described by a stabilizer Hamiltonian $H_{S_i}$, immersed in a thermal bath of absolute temperature $T$, where none of the systems interact with each other. The system Hamiltonian here is given by $H_S=\sum_{i=1}^LH_{S_i}$, while the state of the system is $\rho_S=\otimes_{i=1}^L\rho_{S_i}$, where 
$\rho_{S_i} =\exp\left(-\beta H_{S_i}\right)/Z_i$ $\forall i$ with  $Z_i=\text{Tr}\left[\exp\left(-\beta H_{S_i}\right)\right]$ being the partition function, $\beta=1/k_BT$, and \(k_B\) the Boltzmann constant. Our aim is to \emph{perfectly} purify the thermal state $\rho_{S_i}$ of each QEC to an arbitrary logical state $\ket{\Psi_{S_i}}$. For this,  we bring in a physical \emph{auxiliary} qubit (AQ), $A$, of energy $E_A$, described by the Hamiltonian $H_{A}=(E_A/2)\sigma^z_A$, where $\sigma^z$ is the $z$ component of the Pauli matrices. We further assume that the AQ is fully accessible such that (a) $E_A$ can be set to any value, (b) all possible single-qubit measurements can be performed on it, and (c) it can be initialized to any arbitrary pure state $\ket{\psi_A}=\cos (a/2)\ket{0_A}+\text{e}^{\text{i}b}\sin (a/2) \ket{1_A}$ ($0\leq a\leq \pi$, $0\leq a\leq 2\pi$). Preparing the AQ in its ground state $\ket{0_A}$, the entire QECCs-AQ duo is  initialized to $\rho_{SA}=\left[\otimes_{i=1}^L\rho_{S_i}\right]\otimes \ket{0_A}\bra{0_A}$.  In the following, we demonstrate how the transformation $\rho_{S_i}\rightarrow\ket{\Psi_{S_i}}$ occurs \emph{simultaneously with unit fidelity}.

\noindent\textbf{Proposition.} \emph{An arbitrary thermal state $\rho_S=\otimes_{i=1}^L\rho_{S_i}$ of $L$ QECs can be perfectly purified to $\ket{\Psi_S}=\otimes_{i=1}^L\ket{\Psi_{S_i}}$ with unit fidelity and a finite probability via a time evolution generated by an appropriately engineered Hamiltonian describing interaction between the QECCs and the AQ, followed by a projective measurement on the latter.}

\begin{figure}
    \centering
    \includegraphics[width=\linewidth]{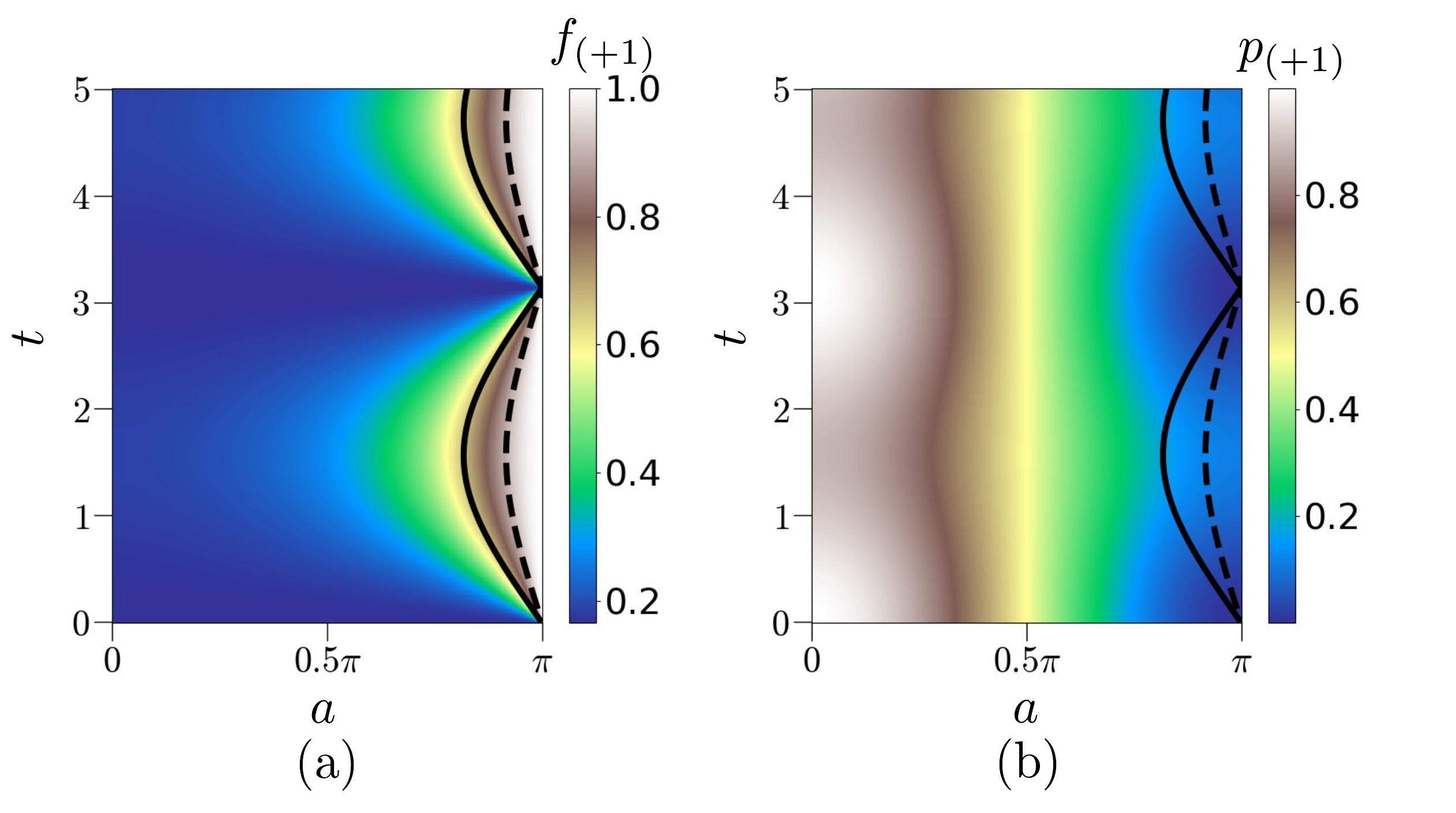}
    \caption{Variations of $f_{(+1)}$ (Eq.~(\ref{eq:fidelity_mes_main_text})) and $p_{(+1)}$ (Eq.~(\ref{eq:probability_main_text})) as a function of $a$ (abscissa) and $t$ (ordinate), with $L=1$, $d=6$, $g=1$, $D=8$, $\Delta E=4$, $E_A=\Delta E$, and $\beta=10^{-1}$. The continuous (dashed) line corresponds to $f_{(+1)}=0.66$ ($f_{(+1)}=0.9$) in the case of both $f_{(+1)}$ (a) and $p_{(+1)}$ (b). All quantities plotted are dimensionless except the axis $a$, which is in radian.}
    \label{fig:fidelity}
\end{figure}

\begin{proof}
Consider an interaction Hamiltonian 

\begin{eqnarray}
    H_{SA}&=&g\ket{\Psi_{S}}\bra{\Phi_{S}}\otimes\ket{1_A}\bra{0_A}+\text{h.c.}  
\label{eq:interaction_degenerate_multiple_LQ}
\end{eqnarray}
of strength $g$, turned on between the state 
\begin{eqnarray}
    \ket{\Psi_{S}}=\bigotimes_{i=1}^L\ket{\Psi_{S_i}}
    \label{eq:ls_state}
\end{eqnarray}
in the LS, and the state 
\begin{eqnarray}
\ket{\Phi_{S}}=\bigotimes_{i=1}^L\frac{1}{\sqrt{d_i}}\sum_{\mu_{S_i}=2}^{d_i+1}\ket{\mu_{S_i}}
\label{eq:es_state}
\end{eqnarray}
in the ES of the QECCs, corresponding to a total energy cost $|E_A-\sum_{i=1}^L\Delta E_i|$ via the AQ $A$ (see Fig.~\ref{fig:schematics}), taking the QECCs-AQ duo out of equilibrium. Starting from the initial state $\rho_{SA}$, the time-evolution is given by $\rho_{SA}(t)=U_{SA}\rho_{SA}U_{SA}^\dagger$, where $U_{SA}=\exp\left[-\text{i}H_{\text{tot}}t\right]$, with $H_{\text{tot}}=H_S+H_A+H_{SA}$. Expanding, one gets  
\begin{eqnarray}
    \rho_{SA}(t)&=& \mathcal{O}_S^1(t)\otimes\ket{1_A}\bra{1_A} + \mathcal{O}_S^0(t)\otimes \ket{0_A}\bra{0_A}\nonumber\\&&+(\mathcal{O}_{S}^{10}(t)\otimes \ket{1_A}\bra{0_A}+\text{h.c.}),
    \label{eq:time_evolution_main_text}
\end{eqnarray}
where (see Appendix~\ref{app:evolution_details} for details)
\begin{eqnarray}
\mathcal{O}_S^1(t)&=&p_{(+1)}\bigotimes_{i=1}^{L}\ket{\Psi_{S_i}}\bra{\Psi_{S_i}},
\end{eqnarray}
and we refrain from including the forms of $\mathcal{O}_S^0(t)$ and $\mathcal{O}_S^{10}(t)$ as they are not relevant for the proof (see Appendix~\ref{app:evolution_details}). Here, 
\begin{eqnarray}
    p_{(+1)}
    &=& 4g^2 p_\beta \left( \frac{E_+^2}{G_{+}^2}+\frac{E_-^2}{G_{-}^2}+\frac{2 E_+ E_- \cos{(Ft)}}{G_{+}G_{-}} \right), 
\end{eqnarray}
with 
\begin{eqnarray}
    p_\beta=\prod_{i=1}^LZ_i^{-1}\exp(-\beta \Delta E_i),
    \label{eq:pbeta_total}
\end{eqnarray}
and $F=\sqrt{(\sum_{i=1}^L\Delta E_i-E_{A})^2+4g^2}$, $E
_{\pm}=\sum_{i=1}^L\Delta E_i-E_A\pm F$, $G_\pm=E_{\pm}^2+4g^2$. At any arbitrary time $t$, a $\sigma^z$-measurement ($\sigma^z$ being the $z$-component of Pauli operators) on the AQ  yields the outcome $+1$ with  probability $p_{(+1)}=\langle 1_A|\rho_{SA}(t)|1_A\rangle$  
which simultaneously results in the post-measured states $\ket{1_A}$ and $\ket{\Psi_S}$ on the AQ and $S$ respectively. Hence the proof. 
\end{proof}

It is worthwhile to question the uniqueness of the choice of $\sigma^z$-measurement on the AQ for preparing $\ket{\Psi_S}$. We pursue this by parametrizing the measurement element on the AQ using two real parameters $a,b$ ($a\in[0,\pi]$, $b\in[0,2\pi]$) as $M_A(k,a,b)=P_A^{(k)\dagger}P_A^{(k)}$ (see Appendix~\ref{app:measurement_parameters}) with $P_A^{(k)}=\ket{\psi_A^{(k)}}\bra{\psi_A^{(k)}}$, and 
\begin{eqnarray}
    \ket{\psi_A^{(+1)}} &=& \cos\frac{a}{2}\ket{0_A}+\text{e}^{\text{i}b}\sin\frac{a}{2}\ket{1_A},\nonumber\\
    \ket{\psi_A^{(-1)}} &=& \sin\frac{a}{2}\ket{0_A}-\text{e}^{\text{i}b}\cos\frac{a}{2}\ket{1_A},
    \label{eq:measurement-basis_main_text}
\end{eqnarray}
where $k=\pm 1$ are the measurement outcomes, occurring respectively with probabilities $p_{(\pm 1)}$, such that $\sum_{k=\pm 1}M_A(k,a,b)=\mathbb{I}$. Note that the outcomes $k=\pm 1$ of the $\sigma^z$-measurement corresponds to the special cases of $a=\pi$ and $a=0$ respectively, irrespective of the irrelevant global phase $\text{e}^{\text{i}b}$. For $E_A=\sum_{i=1}^L\Delta E_i$ implying transitions with no energy cost, the measurement yields the outcome $+1$ with probability 
\begin{eqnarray}
p_{(+1)}&=&p_\beta\Big[\sin^2\left(gt\right) \sin^2\left(\frac{a}{2}\right)+\cos^2\left(gt\right) \cos^2\left(\frac{a}{2}\right)\Big]\nonumber \\
 &&+(1-p_{\beta})\cos^2\left(\frac{a}{2}\right),
\label{eq:probability_main_text}
\end{eqnarray}
where $p_\beta$ is as defined in Eq.~(\ref{eq:pbeta_total}). 
This corresponds to the post-measured state $\rho_{S}^{(+1)}=\text{Tr}_A\left[\rho_{SA}^{(+1)}\right]$ on the QECCs with a fidelity $f_{(+1)}=\langle \Psi_S|\rho_S^{(+1)}|\Psi_S\rangle$, given by  
\begin{eqnarray}
    f_{(+1)}=p_{(+1)}^{-1}\left[Z_L^{-1}\cos^2 \left(\frac{a}{2}\right)+p_\beta\sin^2 \left(gt\right)\sin^2\left(\frac{a}{2}\right)\right],\nonumber\\ 
    \label{eq:fidelity_mes_main_text}
\end{eqnarray} 
where $Z_L=\prod_{i=1}^LZ_i$.

For a specific measurement (i.e., for a constant $a$), $f_{(+1)}$ is maximum when $t=t_n^{\text{max}}=\left(n+\frac{1}{2}\right)\left[g^{-1}\pi\right]$. Note that $a=\pi$ sets $f_{(+1)}=1$ $\forall t$,  implying \emph{perfect purification} of all logical qubits, in  which case, 
\begin{eqnarray}
    p_{(+1)}&=&\frac{p_\beta}{2}\left(1-\cos{(2gt)}\right),
\label{eq:postselection_probability_single_system}
\end{eqnarray} 
is non-vanishing except at\footnote{Note that the probability vanishes also for the trivial case of $g=0$.}
$t=t_n^0=ng^{-1}\pi$,  and is maximum at $t=t_n^{\text{max}}$, where $n=0,1,\cdots$. The maximum  value $p_{(+1)}^{\text{max}}=p_\beta$ of the probability is a function of the initial temperature of the system, and attains the maximum value $p_{(+1)}^{\text{max}}=D^{-L}$ for $\beta=0$, corresponding to the maximally mixed initial states of all $S_i$, $i=1,2,\cdots,L$.

Note that the cost of increasing the number, $L$, of QECCs is reflected in a decrease in the  achievable maximum probability $p_{(+1)}^{\text{max}}$. Further, given a desired value  $\tilde{f}$ of $f_{(+1)}$, the required measurement is fixed by a  specific value of $a$, which can be obtained from Eq.~(\ref{eq:fidelity_mes_main_text}) for specific QECCs. For example, in the case of \emph{identical} ($H_{S_i}=H_S$, $D_i=D$ $\forall i$) QECCs with $d_i=d=D-2$\footnote{This corresponds to the three-qubit repetition code. See Eq.~(\ref{eq:three_qubit_code}) and the corresponding discussions.} $\forall L$,  
\begin{equation}
a=2\cos^{-1}\left[B^{-1/2}\sqrt{\left(1-\tilde{f}\right)\left(1-\cos2gt\right)}\right],
\label{eq:mes_setting}
\end{equation}
while the \emph{physical} space of the parameters $g,t,\beta$, and $\Delta E_i$ is determined from the positivity of
\begin{eqnarray}
    B&=&\nonumber 2\exp\left(\beta\sum_{i=1}^L\Delta E_i\right)Z_L\tilde{f}-2\Bigg [\exp\left(\beta\sum_{i=1}^L\Delta E_i\right)\\&& -\left(1-2\tilde{f}\right)\sin^2gt\Bigg ].\nonumber\\ 
\end{eqnarray}
Fig.~\ref{fig:fidelity} depicts the variations of $p_{(+1)}$  and $f_{(+1)}$ as functions of $a$ and $t$ corresponding to a QECC ($L=1$) with $d=D-2$. It is clear from the figure that for the specific point $(g,\beta,\Delta E)$ in the parameter space, there may exist measurements such that a fidelity $f_{(+1)}>0.66$ -- the \emph{classical limit} for the fidelity~\cite{MassarPopescu95} -- can always be achieved with non-zero probability. Further, for the same point in the parameter space, it may also be possible to find a measurement to obtain an arbitrary logical state with $f_{(+1)}>0.9$ with $p_{(+1)}>0$. This implies that even when the measurement choice of $\sigma^z$ is relaxed, the proposed scheme can be successful with a high fidelity and a finite probability, thereby establishing the flexibility of the scheme over a large set of measurements.

\noindent\textbf{Note.} It is important to point out the following subtleties regarding the interaction Hamiltonian. Instead of Eq. (\ref{eq:es_state}), one can also consider $\ket{\Phi_S}$ to be an arbitrary superposition of the ES states as 
        \(\ket{\Phi_S}=\bigotimes_{i=1}^L\sum_{\mu_{S_i}=2}^{d_i+1}\alpha_{\mu_{S_i}}\ket{\mu_{S_i}}\),
with $\sum_{\mu_{S_i}=2}^{d_i-1}|\alpha_{\mu_{S_i}}|=1$, without any change in the results. Further, 
a more general variant of the interaction Hamiltonian having the form 
    \begin{eqnarray}
H_{SA}&=&g\left[\bigotimes_{i=1}^L\left(\sum_{\mu_{S_i}=2}^{d_i+1}\ket{\Psi_{S_i}}\bra{\mu_{S_i}}\right)\right]\otimes\ket{1_A}\bra{0_A}+\text{h.c.}, \nonumber\\ 
    \label{eq:interaction_targeted_LQ}
    \end{eqnarray}
qualitatively leads to the same results, except the expressions for $f_{(+1)}$ and $p_{(+1)}$, which, in general, become more cumbersome.

\paragraph{Demonstration with three-qubit repetion code.} Typically, each QECC $S_i$ is made of $N_i$ physical qubits, referred to as the \emph{data qubits} (DQs), such that $D_i=2^{N_i}$, where each $H_{S_i}$ represents a stabilizer Hamiltonian.
In a stabilizer QEC, a set of \emph{stabilizer} operators $\boldsymbol{\mathcal{S}}=\{\mathcal{S}_i\}$, constructing an Abelian subgroup of the Pauli group, \emph{stabilize}  all logical states $\ket{\Psi_S}$ of the code, i.e., $\mathcal{S}_i\ket{\Psi_S}=(+1)\ket{\Psi_S}$ $\forall i$ and $\forall$ $\ket{\Psi_S}$, along with satisfying $[\mathcal{S}_i,\mathcal{S}_j]=0$ $\forall \mathcal{S}_i,\mathcal{S}_j\in\boldsymbol{\mathcal{S}}$. One can interpret the logical subspace as the ground state subspace of the stabilizer Hamiltonian $H_S=-\sum_{\mathcal{S}_i\in\boldsymbol{\mathcal{S}}}\mathcal{S}_i+|\boldsymbol{\mathcal{S}}|I$, where $|\boldsymbol{\mathcal{S}}|$ is the cardinality of $\boldsymbol{\mathcal{S}}$, and $I$ is the identity operator on the Hilbert space of the system. Note that the term $|\boldsymbol{\mathcal{S}}|I$ ensures that the ground state energy of $H_S$ vanishes.

\begin{figure}
\includegraphics[width=0.75\linewidth]{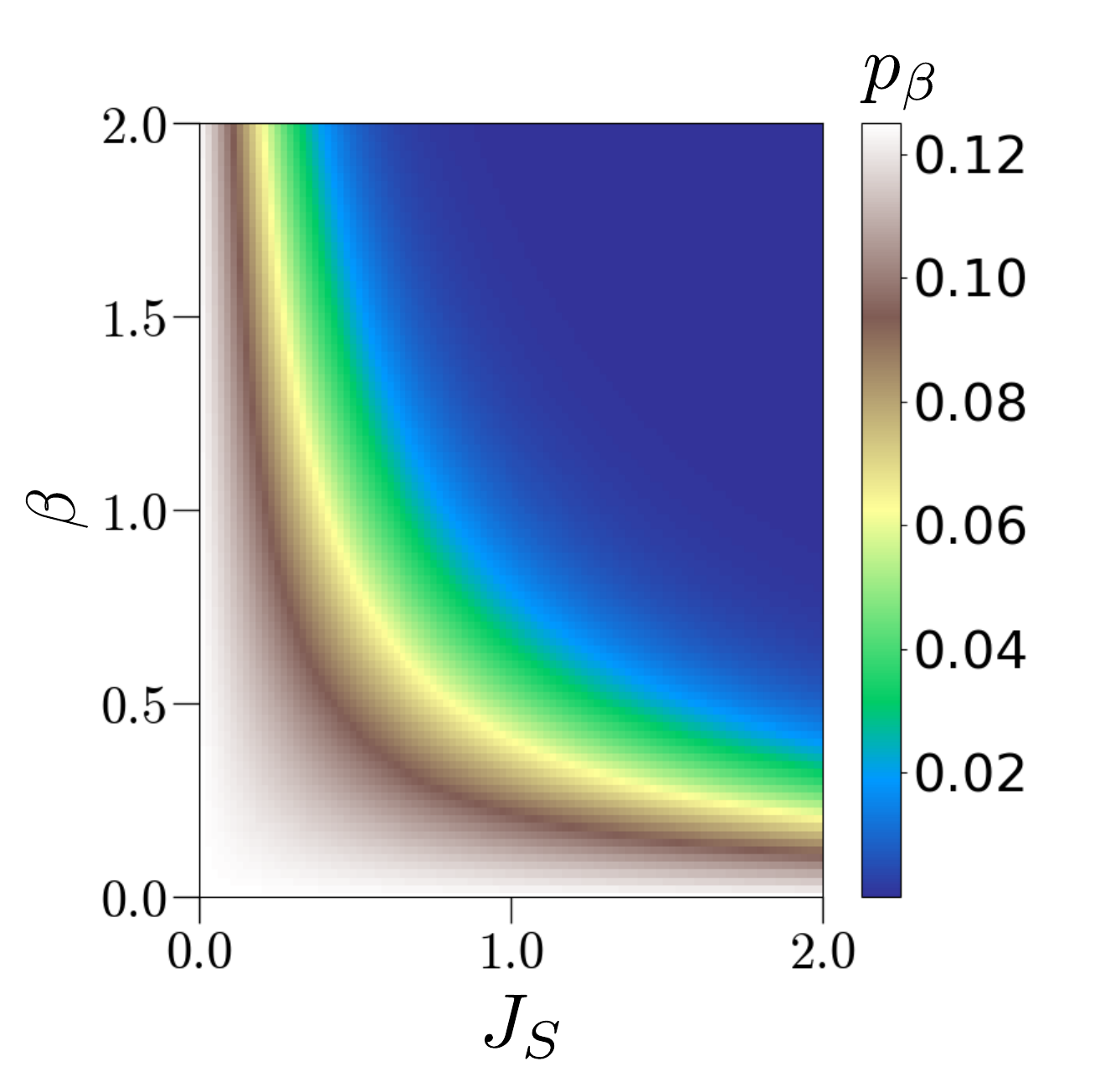}
\caption{Map plot of $p_\beta$ with respect to  $J_S$ (abscissa), and $\beta$ (ordinate) for the three-qubit bit-flip code. All quantities plotted are dimensionless.}
\label{fig:pbeta_stabilizer}
\end{figure}

A specific example is a $3$-qubit $(D=8)$ repetition code defined by the stabilizers
\begin{eqnarray}
    \mathcal{S}_1=\sigma^z_1\sigma^z_2, \;
    \mathcal{S}_2=\sigma^z_2\sigma^z_3, \; \mathcal{S}_3=\sigma^z_3\sigma^z_1,
    \label{eq:three_qubit_code}
\end{eqnarray}
where the logical subspace is given by the two-fold degenerate ground state subspace of the Hamiltonian $H_S=-J_S\sum_{i=1}^3\mathcal{S}_i+3J_SI$. The logical $Z$ basis in this case is given by $\ket{0_S}=\ket{000},\ket{1_S}=\ket{111}$ spanning the logical subspace, which is separated from the six-fold degenerate ($d=6$) first excited state by an energy gap $\Delta E=4J_S$. The interaction Hamiltonian $H_{SA}$ corresponding to $\ket{\Psi_S}$ for the three-qubit repetition code ($L=1$) reads 
\begin{widetext}
\begin{eqnarray}
    H_{SA}&=&\frac{g}{8\sqrt{6}}\sum_{j=1}^{3}\Big[(\zeta_+ \sigma^x_A-\zeta_{-}\sigma^z_j \sigma^x_A)(\sigma^x_{j+1}\sigma^x_{j+2}-\sigma^y_{j+1}\sigma^y_{j+2})+(\zeta_- \sigma^y_A-\zeta_{+}\sigma^z_j \sigma^y_A)(\sigma^x_{j+1}\sigma^y_{j+2}+\sigma^y_{j+1}\sigma^x_{j+2})+\zeta_+(\sigma^x_j\sigma^x_A \nonumber \\&-&\sigma^y_j\sigma^z_{j+1}\sigma^y_A-\sigma^y_j\sigma^z_{j+2}\sigma^y_A+\sigma^x_j\sigma^z_{j+1}\sigma^z_{j+2}\sigma^x_A) +\zeta_-
    (\sigma^y_j\sigma^y_A-\sigma^x_j\sigma^z_{j+1}\sigma^x_A-\sigma^x_j\sigma^z_{j+2}\sigma^x_A +\sigma^y_j\sigma^z_{j+1}\sigma^z_{j+2}\sigma^y_A)\nonumber \\
    &-&\sin{\left(\frac{\theta}{2}\right)}\sin{\phi}\Big\{(\mathbb{I}+\sigma^z)_j(\sigma^x_{j+1}\sigma^x_{j+2}\sigma^y_{A} +\sigma^x_{j+1}\sigma^y_{j+2}\sigma^x_{A}+\sigma^y_{j+1}\sigma^x_{j+2}\sigma^x_{A}-\sigma^y_{j+1}\sigma^y_{j+2}\sigma^y_{A})+(\sigma^x_j \sigma^y_{A}\nonumber \\
    &+&\sigma^y_j \sigma^x_{A})(\mathbb{I}+\sigma^z_{j+1}+\sigma^z_{j+2}+\sigma^z_{j+1}\sigma^z_{j+2})\Big\}\Big] 
\end{eqnarray}
\end{widetext}
in the \emph{canonical form} in terms of the Pauli operators for the DQs in the QECC,  with  $\zeta_\pm=\cos{\theta/2}\pm\cos{\phi}\sin{\theta/2}$ ($0\leq\theta\leq\pi$,  $0\leq\phi\leq 2\pi$). The involvement of only two- and few-body interactions, and the fact that these interactions are implementable in trapped ions~\cite{Mintert2001,*porras2004} and superconducting qubits~\cite{dalmonte2015} using recent technologies, make the realization of the proposed protocol possible. Note, however, that determination of the canonical form of $H_{SA}$ requires access to ES, and therefore is difficult for QECCs with large distances. For $L=1$, with $\sigma^z$-measurement, the variation of $p_{(+1)}^{\text{max}}=p_\beta$ (see discussion following Eq.~(\ref{eq:postselection_probability_single_system})) as a function of $J_S$ and $\beta$ in the case of the three-qubit repetition code is given
in Fig.~\ref{fig:pbeta_stabilizer}. Note that $f_{(+1)}$ remains independent of the values of $t$, $g$, and the choice of the QECC, while $p_{(+1)}^{\text{max}}$ varies with the size $N$ of the chosen QECC as $2^{-N}$ for $L=1$.

\paragraph{Purification of logical qubits hosted in Heisenberg model.} An arbitrary logical state corresponding to the logical qubit required for quantum state transfer in 2D lattices~\cite{Pushpan2024Jul} can also  be prepared using the same protocol. The logical qubit is hosted in the doubly-degenerate ground state subspace of a system on $N$ DQs on a 1D lattice with periodic boundary condition, described by the isotropic Heisenberg Hamiltonian~\cite{fisher1964,*Mila_2000,*giamarchi2004,*Franchini2017} 
\begin{eqnarray}
    H_S=\frac{J_S}{4}\sum_{s=1}^{N}\vec{\sigma}_s.\vec{\sigma}_{s+1}+\frac{h_S}{2}\sum_{s=1}^N\sigma^z_s-E_gI,
    \label{eq:N_qubit_system_hamiltonian}
\end{eqnarray}
where $\vec{\sigma}=\left(\sigma^x,\sigma^y,\sigma^z\right)$, $J_S$ is the strength of the exchange interaction between the qubits labeled by $s=1,2,\cdots,N$, $h_S$ is the local magnetic field strength fixed at $h_S=2J_S$ ($h_S=J_S$) for $N>2$ ($N=2$), and we assume $N$ to be even~\cite{Pushpan2024}. The constant $E_g$,  given by 
\begin{eqnarray}
    E_g&=&\left\{\begin{array}{c}
         \left(\frac{J_S}{4}+h_S\right)\text{ for }N=2,\\
        N\left[\frac{1}{4}+\frac{h_S}{2J_S}\right] \text{ for }N>2,
    \end{array}\right.
\end{eqnarray}
The logical $Z$ basis spanning the LS is given by 
\begin{eqnarray}
    \ket{0_S}&=&\ket{0}^{\otimes N},\nonumber\\ 
    \ket{1_S}&=&\frac{1}{\sqrt{N}}\sum_{s=1}^N(-1)^s\sigma^x_s\ket{0}^{\otimes N},
    \label{eq:heisenberg_logical_state}
\end{eqnarray}
which are separated from the ES by a gap $\Delta E\sim J_S$, implying that the variation of $p_\beta$ remains qualitatively the same as the three-qubit repetition code (see Fig.~\ref{fig:pbeta_stabilizer}). However, unlike the latter case where $p_\beta$ remains constant with increasing $N$ by choosing QECCs with larger distances, $p_\beta\sim 2^{-N}$ in the present case.

\subsection{Single measurement-based protocol for ground state preparation}
\label{subsec:ground_state_preparation}

A similar protocol can also be designed for preparing the non-degenerate ground states of $L$ arbitrary quantum systems (not necessarily a QECCs). Consider the case of the $i$th system,  where the ground state $\ket{0_{S_i}}$ has the energy $E_0=0$, and the first excited state with energy $E_1$ $(\Delta E_i=E_1)$  is $d_i$-fold ($d_i\geq 1$) degenerate. Using an AQ initialized in the ground state $\ket{0_A}$, and an interaction Hamiltonian of the form
\begin{eqnarray}
    H_{SA}&=&g\ket{0_S}\bra{\Phi_S}\otimes\ket{1_A}\bra{0_A}+\text{h.c.},
\label{eq:interaction_non_degenerate_multiple}
\end{eqnarray}
with $\ket{0_S}=\bigotimes_{i=1}^L\ket{0_{S_i}}$, and
\begin{eqnarray}
    \ket{\Phi_S}=\bigotimes_{i=1}^L\frac{1}{\sqrt{d_i}}\sum_{\mu_{S_i}=1}^{d_i}\ket{\mu_{S_i}},
\end{eqnarray}
the above protocol works as a ground state preparation protocol for each of the system, as described in the following Corollary.     

\noindent\textbf{Corollary.} \emph{The ground states $\ket{0_{S_i}}$s of each of $L$ quantum systems $S_i,i=1,2,\cdots, L$, can be simultaneously and perfectly purified with a finite probability, starting from an arbitrary thermal state of the form  $\rho_S=\otimes_{i=1}^L\rho_{S_i}$ via a time evolution generated by $H_{SA}$ (in Eq.~(\ref{eq:interaction_non_degenerate_multiple})), followed by a single projective measurement on the AQ.}

\noindent The proof is similar to that of the \textbf{Proposition} (see Appendix~\ref{app:evolution_details}).

 


\begin{figure}
\includegraphics[width=\linewidth]{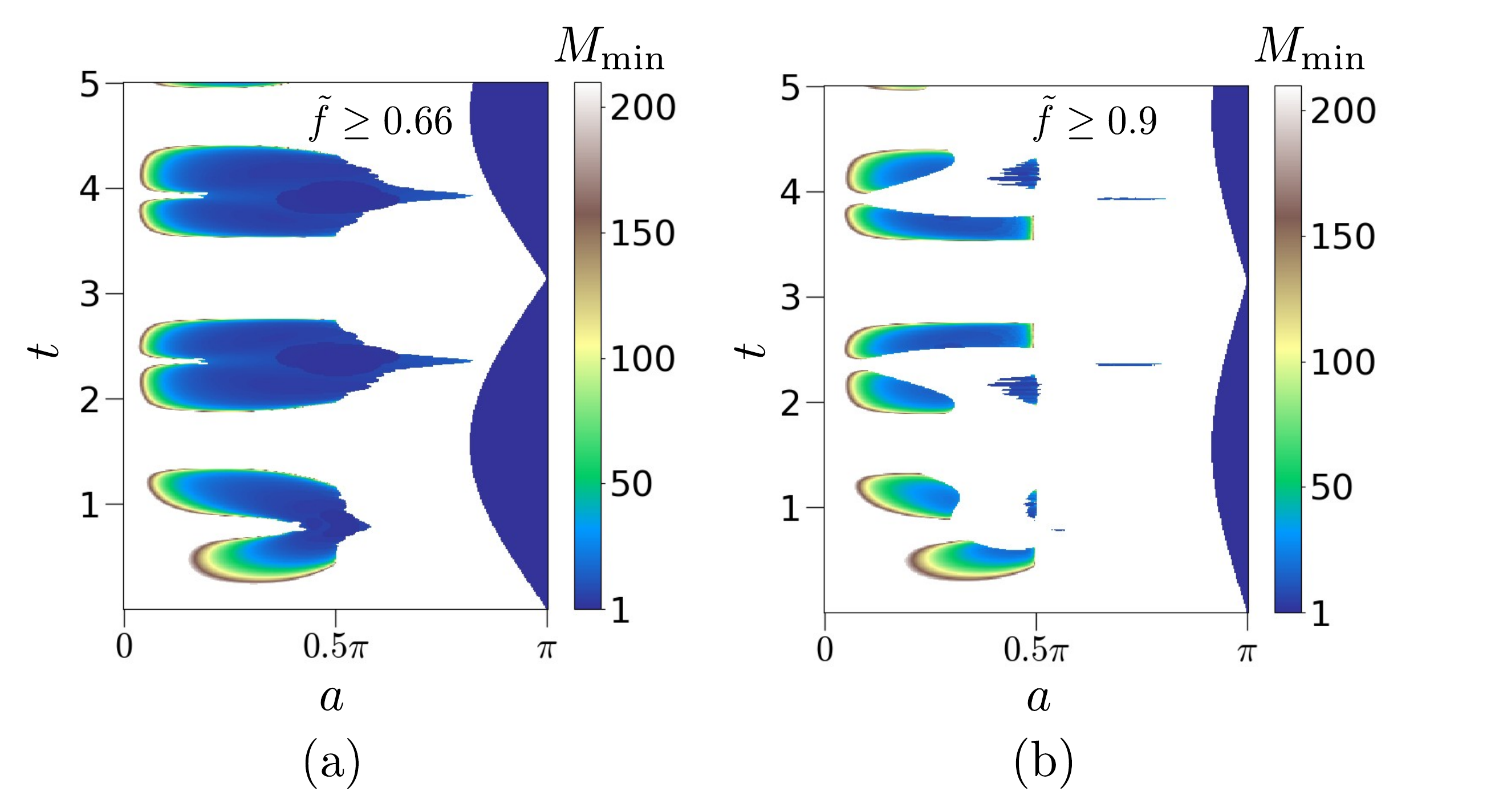}
\caption{$M_{\text{min}}$ against $a$ (horizontal axis) and $t$ (vertical axis) for (a) $\tilde{f}=0.66$ and (b) $\tilde{f}=0.9$ in the case of the three-qubit repetition code ($D=8$, $d=6$, $J_S=1$), where we fix $g=1$, $E_A=\Delta E=4$, and $\beta=10^{-1}$. Each measurement (i.e., each value of $a$) is repeated for up to $200$ rounds. The data indicates that majority of the points $(a,t)$ corresponds to measurements and evolution times for which the desired fidelity $\tilde{f}$ is not achieved even after 200 rounds of measurements.  All quantities plotted are dimensionless except the $a$ axis, which is in radian.}
\label{fig:multiple_measurements}
\end{figure}

\section{Improving sub-optimal fidelity with repeated purification rounds}
\label{sec:repeated_measurements}

Upto now, a time-evolution of the QECCs-AQ duo via the engineered $H_{SA}$ (in Eq. \ref{eq:interaction_degenerate_multiple_LQ}) followed by a single $\sigma^z$-measurement of the AQ and corresponding post-selection has led to a perfect purification.  
However, when the fidelity of the purified logical state is sub-optimal for a specific value of $a$ corresponding to a single purification round (see Fig.~\ref{fig:fidelity}),  it is logical to explore extensions of the purification protocol such that a fidelity $\geq \tilde{f}$, where $\tilde{f}$ is a desired value of $f_{(+1)}$, can be achieved. To this end, for fixed values of $a$ and $\tilde{f}$ where one round of purification does not provide the desired fidelity but completely disentangles the QECCs and the AQ, we ask the following question: \emph{Can repeated rounds of purification enhance the sub-optimal fidelity obtained in a single purification round?} This question is relevant in situations where one, while being forced to work with a specific QECC and at a given point $(g,\Delta E,\beta)$ in the parameter space, can still perform multiple rounds of the same measurements (i.e., a fixed value of $(a,b)$) on the fully accessible AQ. Our investigation answers this question affirmatively. For brevity, in what follows, we denote \emph{one round of purification} by $\{t,(a,b),k\}$, depending on the specific values of $t$, $(a,b)$, and $k$. We call the protocol involving repeated rounds of purification by the name ``\emph{evolve-measure-repeat}" (EMR) protocol.

We assume that in each round of purification in the EMR protocol, the outcome $(+1)$ is chosen (i.e., $k=+1\forall r$), and the resulting fidelity $f_{(+1)}$ corresponds to a string of measurement outcomes $(+1)_1(+1)_2\cdots(+1)_r\cdots(+1)_M$ in $M$ rounds, where each round correspond to a measurement fixed by $(a,b)$.  
Our analysis reveals that for a considerable set of measurements and interactions between the QECCs and the AQ, the classical fidelity as well as a fidelity more than $0.9$ of the logical state can be attained for a finite number of rounds of purification scheme, denoted by $M_{\text{min}}$ (see Fig.~\ref{fig:multiple_measurements}). 
Note further that one can also post select different outcomes in each rounds of purification, and there are $2^M$ such possible sets of outcomes. While we restrict ourselves to the specific case of $(+1)$ outcome post each purification round, the absolute minimum number of rounds required to achieve a desired $\tilde{f}$ can be determined via a minimization over all possible sets of measurement outcomes, which is difficult owing to the exponential growth of the set with $M$. However, our analysis with $k=(+1)$ provides at least an upper bound of $M_{\text{min}}$, if not the exact value.

\emph{Can the power of EMR protocol be used to achieve purification of the logical states with high fidelity using interaction Hamiltonians that are not of the form of $H_{SA}$ (Eq.~(\ref{eq:interaction_degenerate_multiple_LQ}))}? To answer this, we consider the example of the two-qubit isotropic Heisenberg Hamiltonian as $H_S$, and define the interaction between the system and the AQ by~\cite{Barouch1970,*Barouch1971,*Barouch1971a,Sachdev2011}    
\begin{eqnarray}
    H_{SA}=\sum_{i=1,2}\frac{J_i}{4}\left(\gamma_+\sigma^x_A\sigma^x_i+\gamma_-\sigma^y_A\sigma^y_i\right),
    \label{eq:xy-interaction}
\end{eqnarray}
where $J_{i}$ is the strength of the exchange interaction between the ancillary DQ $A$ and a system-DQ $i$ ($i=1,2$), and $\gamma_\pm=(1\pm\gamma)/2$ with $\gamma$  being the $xy$-anisotropy parameter ($-1\leq\gamma\leq 1$). Setting $J_1=1$, $J_2=0$, and $\gamma=0$, $M<15$ repeated rounds given by $\{1,(0,0),+1\}$ prepares the logical states $\ket{0_S}$ and $\ket{1_S}$ with fidelity $f_{(+1)}=1$. The probability $p_{(+1)}=\prod_{r=1}^{M}p_{(+1)}^{(r)}$, on the other hand, decreases with increasing $r$, and saturates to $p_{(+1)}=0.268$. However,  achieving all states in the logical subspace of the two-qubit isotropic Heisenberg Hamiltonian is not possible using the $\sigma^z$ measurement alone, and the choice of the measurement outcome $k$ in each round also plays an important role in the performance of the EMR protocol. Further, with increasing $N$,  our analysis indicates the requirement of multiple auxiliary qubits to achieve a given logical state with a good fidelity, although the minimum number of auxiliary qubits $\ll N/2$ (see Appendix~\ref{app:xy_interaction}).  However, it is noteworthy that interactions similar to that given in Eq.~(\ref{eq:xy-interaction}) is inadequate to purify the logical states of a QECC code, highlighting the importance of the proposed Hamiltonian (Eq.~(\ref{eq:interaction_degenerate_multiple_LQ})).

\section{Conclusion}
\label{sec:conclusion}


We designed a measurement-based protocol for perfectly purifying logical states of arbitrary quantum error correcting codes (QECCs) from their arbitrary thermal states using an auxiliary qubit. It is achieved via a time-evolution generated by an engineered Hamiltonian involving the QECCs and the auxiliary qubit, followed by a single measurement on the auxiliary qubit and an appropriate post-selection of one of the measurement outcomes. The interaction Hamiltonian involves the target logical state(s) on the QECCs as well as the lowest energy states of the error subspaces. We proved that the fidelity of obtaining the desired logical state using this protocol is unity irrespective of the duration of the time evolution, the strength of the interaction Hamiltonian, and the temperature of the QECCs at initial time. We further showed that the probability of obtaining the desired logical state(s) is finite, underscoring the practical utility of the approach. 

Towards establishing the robustness of the protocol, we demonstrated that the other measurements on the auxiliary qubit can not only surpass the classical fidelity, but can reach a fidelity as high as $90\%$. Furthermore, we found that the sub-optimal fidelity obtained in a single round of the purification scheme can be overcome by multiple rounds. Other than QECCs, we also explored the applicability of the protocol in perfectly purifying the logical states useful for quantum state transfer emphasizing its universality, and for using paradigmatic quantum spin Hamiltonians to generate the necessary time evolution for perfect purification. The feasibility of implementing the spin-spin interactions involved in the canonical form of the engineered interaction Hamiltonian using current technology makes the experimental realization of the proposed scheme a possibility. 

The proposed scheme employs a minimal number of auxiliary qubits and measurements to accomplish the goal of purifying logical qubits, and provides a new avenue to generate resource for QECCs, especially when the codes are exposed to thermal noise. The effect of other noise sources on the logical qubits as well as the presence of imperfections in the engineered Hamiltonian on the developed protocol remain to be explored.  


\acknowledgements

The authors acknowledge the cluster computing facility at Harish-Chandra Research Institute and the use of \href{https://github.com/titaschanda/QIClib}{QIClib} -- a modern C++ library for general purpose quantum information processing and quantum computing. This research was supported in part by the ``INFOSYS scholarship for senior students''.  A.K.P and A.S.D acknowledge the support from the Anusandhan National Research Foundation (ANRF) of the Department of Science and Technology (DST), India, through the Core Research Grant (CRG) (File No. CRG/2023/001217, Sanction Date 16 May 2024).

\onecolumngrid   

\appendix

\section{Purification of logical states}
\label{app:evolution_details}

In this Section, we provide the necessary details required to follow the time evolution of the QECCs-AQ pair and the following measurement of the AQ. Consider a collection $S$ of $L$ QECCs, labeled by $S_1,S_2,\cdots,S_L$, with Hilbert space dimensions $D_1,D_2,\cdots,D_L$, respectively. The full system is described by the Hamiltonian $H_S=\otimes_{i=1}^LH_{S_i}$, where individual stabilizer Hamiltonians follow $H_{S_i}\ket{\mu_{S_i}}=E_{\mu_i}\ket{\mu_{S_i}}$, $\mu_{S_i}=0,1,\cdots,D_{i}-1$, with a doubly degenerate ground state (see discussions following Eq.~(\ref{eq:general_system_hamiltonian})) constructing the LS. Each of the QECC $S_i$ is initialized in a thermal state $\rho_{S_i}=\exp\left(-\beta_iH_{S_i}\right)/Z_i$ where $Z_i=\text{Tr}\left[\exp\left(-\beta_iH_{S_i}\right)\right]$, and $\beta_i=(k_B T_i)^{-1}$, with $T_i$ being the absolute temperature of $S_i$ and $k_B$ is the Boltzmann constant, such that $\rho_S=\otimes_{i=1}^L\rho_{S_i}$. With the AQ, $A$, in the ground state $\ket{0_A}$ of the Hamiltonian $H_{A}=(E_A/2)\sigma^z_A$, the initial state of the QECCs-AQ duo is given by
\begin{eqnarray}
    \rho_{SA}(0)=\bigotimes_{i=1}^L\left[\sum_{\mu_{S_i}=0}^{D_i-1}\frac{\text{e}^{-\beta_iE_{\mu_{S_i}}}}{\text{Tr}[\exp[-\beta_iH_{S_i}]]} \ket{\mu_{S_i}}\bra{\mu_{S_i}}\right]\otimes \ket{0_A}\bra{0_A}.
\end{eqnarray}
Turning on the QECCs-AQ interaction Hamiltonian of the form (see Eq.~(\ref{eq:interaction_degenerate_multiple_LQ})) 
\begin{eqnarray}
    H_{SA}&=&g\left[\bigotimes_{i=1}^L\left(\frac{1}{\sqrt{d_i}}\sum_{\mu_{S_i}=2}^{d_i+1}\ket{\Psi_{S_i}}\bra{\mu_{S_i}}\right)\right]\otimes\ket{1_A}\bra{0_A} +\text{h.c.}
\end{eqnarray}
results in a time-evolution of the $L$ QECCs and the AQ as $\rho_{SA}(0)\rightarrow\rho_{SA}(t)=U\rho_{SA}U^\dagger$ with $U=\exp\left[-\text{i}H_{\text{tot}}t\right]$ where $H_{\text{tot}}=H_{S}+H_{A}+H_{SA}$, and $d_i$ is the degeneracy of the first excited state $\ket{2_{S_i}}$ of the $i$th system. The eigenvalues of $H_{\text{tot}}$ that are relevant to this calculation are given by 
\begin{eqnarray}
    \ket{\lambda_0}&=&\frac{-2g}{\sqrt{E_{+}^2+4g^2}}\left[\frac{E_{+}}{2g}\bigotimes_{i=1}^L \ket{\Psi_{S_i} 1_A}-\bigotimes_{i=1}^{L}\sum_{\mu_{S_i}=2}^{d_i+1} \frac{1}{\sqrt{d_i}}\ket{\mu_{S_i} 0_A}\right], \\
    \ket{\lambda_1}&=&\frac{-2g}{\sqrt{E_{-}^2+4g^2}}\left[\frac{E_{-}}{2g}\bigotimes_{i=1}^L \ket{\Psi_{S_i}1_A}-\bigotimes_{i=1}^{L}\sum_{\mu_{S_i}=2}^{d_i+1} \frac{1}{\sqrt{d_i}}\ket{\mu_{S_i} 0_A}\right], 
\end{eqnarray}
corresponding to the eigenvalues
\begin{eqnarray}
    \lambda_0&=&\frac{1}{2}\left(E_{A}+\sum_{i=1}^L\Delta E_i-F\right),  \\
    \lambda_1&=&\frac{1}{2}\left(E_{A}+\sum_{i=1}^L\Delta E_i+F\right),  
\end{eqnarray}
respectively, with  
$F=\sqrt{\sum_{i=1}^L\Delta E_i-E_{A}+4g^2}$, $E_{\pm}=\sum_{i=1}^L\Delta E_i-E_A\pm F$. Further, we assume all QECCs to be immersed in the same bath of temperature $T$, such that $\beta_i=\beta$ $\forall i$.   Starting from $\rho_{SA}$, the time-evolved state of the QECCs-AQ duo has the form 
\begin{eqnarray}
    \rho_{SA}(t)&=& \mathcal{O}_S^1(t)\otimes\ket{1_A}\bra{1_A} + \mathcal{O}_S^0(t)\otimes \ket{0_A}\bra{0_A}+(\mathcal{O}_{S}^{10}(t)\otimes \ket{1_A}\bra{0_A}+\text{h.c.}),
    \label{eq:time_evolution}
\end{eqnarray}
with   
\begin{eqnarray}
\mathcal{O}_S^1(t) &=& p_1(t)\left(\bigotimes_{i=1}^{L}\ket{\Psi_{S_i}}\bra{\Psi_{S_i}}\right),\\
\mathcal{O}_{S}^{10}(t)&=&p_{10}(t)\bigotimes_{i=1}^{L}\frac{1}{\sqrt{d_i}}\sum_{\mu_{S_i}=2}^{d_i+1}\ket{\Psi_{S_i}}\bra{\mu_{S_i}},\\ 
\mathcal{O}_S^{0}(t)&=&\rho_{SA}(0) + (p_0(t)-p_\beta)\bigotimes_{i=1}^{L} \frac{1}{d_i}\sum_{\mu_{S_i},\mu_{S_i}^{\prime}=2}^{d_i+1}\ket{\mu_{S_i}}\bra{\mu^{\prime}_{S_i}},
\label{eq:time_evolution_simultaneous}
\end{eqnarray}
where 
\begin{eqnarray}
    p_\beta=\prod_{i=1}^LZ_i^{-1}\exp\left[-\beta\Delta E_i\right].
\end{eqnarray}
In this calculation, we have assumed the ground state energy of all $S_i$, $i=1,2,\cdots,L$ to be zero. The coefficients $p_1(t)$, $p_0(t)$, and $p_{10}(t)$ are given by 
\begin{eqnarray}
    \label{eq:prob_0_L}
    p_0(t)&=&16g^4p_\beta\left(\frac{1}{G_{+}^2}+\frac{1}{G_{-}^2}+\frac{2\cos{(Ft)}}{G_{+}G_{-}}\right),\\
    \label{eq:prob_1_L}
    p_1(t)&=&4 g^2 p_\beta \left( \frac{E_+^2}{G_{+}^2}+\frac{E_-^2}{G_{-}^2}+\frac{2 E_+ E_- \cos{(Ft)}}{G_{+}G_{-}} \right), \\ 
     \label{eq:prob_10_L}
    p_{10}(t)&=&8 g^3 p_\beta \left( \frac{E_+}{G_{+}^2}+\frac{E_-}{G_{-}^2}+\frac{E_+  \text{e}^{iFt}}{G_{+}G_{-}}+\frac{E_-  \text{e}^{-iFt}}{G_{+}G_{-}} \right), 
\end{eqnarray}
where  $G_\pm=E_{\pm}+4g^2$.

An arbitrary projection measurement (see Appendix~\ref{app:measurement_parameters}) on the AQ yields the outcome $+1$ with probability 
\begin{eqnarray}
p_{(+1)}=p_1(t)\sin^2\left(\frac{a}{2}\right)+ p_0(t) \cos^2\left(\frac{a}{2}\right)+\left(1-p_{\beta}\right)\cos^2\left(\frac{a}{2}\right),
\end{eqnarray}
leaving the QECCs-AQ pair in the state 
\begin{eqnarray}
\rho^{(+1)}_{SA}&=& p_1(t)\sin^2 \left(\frac{a}{2}\right)\bigotimes_{i=1}^{L}\ket{\Psi_{S_i}}\bra{\Psi_{S_i}}+\cos^2\left(\frac{a}{2}\right) \Big\{(p_0(t)-p_{\beta})\bigotimes_{i=1}^{L} \frac{1}{d_i}\sum_{\mu_{S_i},\mu_{S_i}^{\prime}=2}^{d_i+1}\ket{\mu_{S_i}}\bra{\mu^{\prime}_{S_i}}+\rho_{SA}(0)\Big\},
\label{eq:post_meausred_state}
\end{eqnarray}
where $\ket{\psi_{(+1)}}$ is given in Eq.~(\ref{eq:measurement-basis}). Fixing $E_A=\sum_{i=1}^L\Delta E_i$, one obtains 
\begin{eqnarray}
    p_{(+1)}=p_\beta\left[\sin^2\left(gt\right) \sin^2\left(\frac{a}{2}\right)+\cos^2\left(gt\right) \cos^2\left(\frac{a}{2}\right)\right]+(1-p_{\beta})\cos^2\left(\frac{a}{2}\right),
\end{eqnarray}
while the post-measured state on the QECCs, $\rho_{S}^{(+1)}=\text{Tr}_A\left[\rho_{SA}^{(+1)}\right]$, has a fidelity $f_{(+1)}=\langle \Psi_S|\rho_S^{(+1)}|\Psi_S\rangle$, given by 
\begin{eqnarray}
    f_{(+1)}=p_{(+1)}^{-1}\left[Z^{-1}_L\cos^2 \left(\frac{a}{2}\right)+p_\beta\sin^2 \left(gt\right)\sin^2\left(\frac{a}{2}\right)\right],
    \label{eq:fidelity_mes}
\end{eqnarray}
where $\ket{\Psi_{S}}=\otimes_{i=1}^L\ket{\Psi_{S_i}}$.

\begin{figure}[H]
    \centering
    \includegraphics[width=0.8\linewidth]{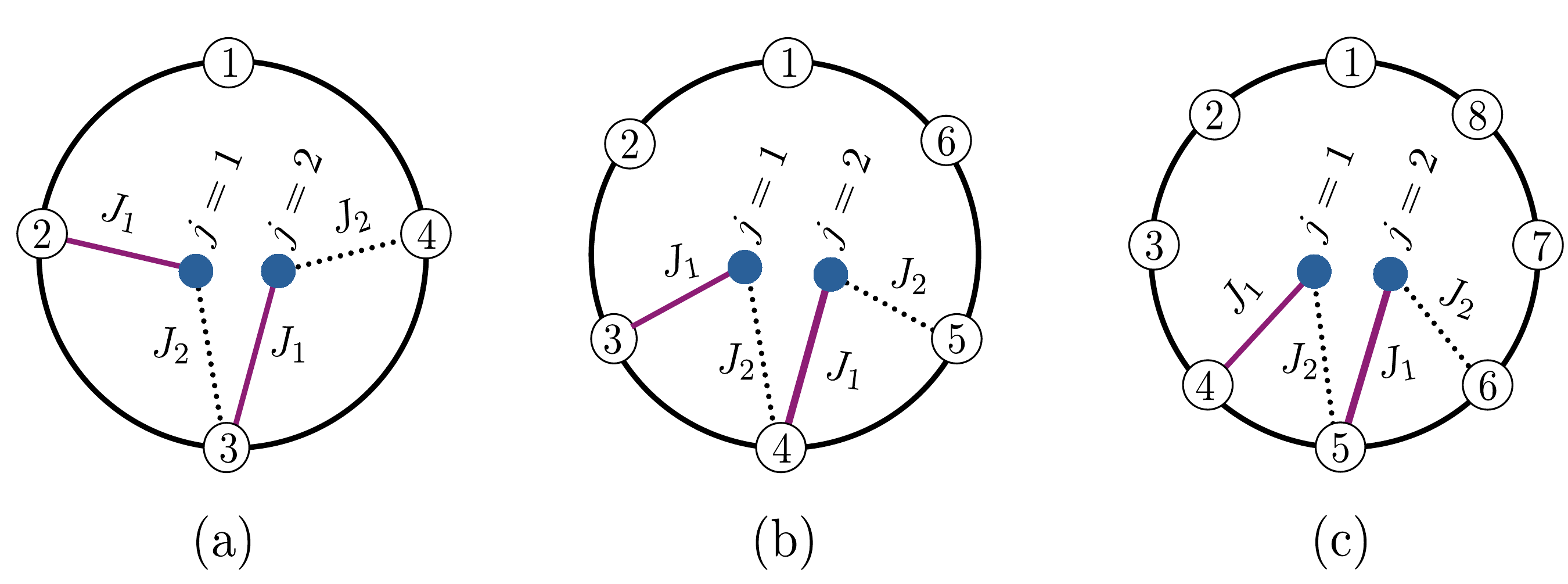}
    \caption{Schematic representation of the $(N,N_A)$-setup with (a) $N=4$, (b) $N=6$, and (c) $N=8$, where $N_A=2$ for all cases.}
    \label{fig:N_qubit_xy}
\end{figure}

\section{Parameterizing single-qubit projection measurements}
\label{app:measurement_parameters}

Consider a projection measurement with measurement element 
\begin{eqnarray}
    M_A(k,\hat{n})=\frac{1}{2}\left(I+k\hat{n}.\vec{\sigma}\right)
\end{eqnarray}
along $\hat{n}$ is performed on a qubit $A$, where  $k=\pm 1$ are the measurement outcomes, $\vec{\sigma}=\left(\sigma^x,\sigma^y,\sigma^z\right)$, and $\hat{n}=\left(n^x,n^y,n^z\right)$ such that $\sum_{\alpha=x,y,z}(n^\alpha)^2=1$
One can parametrize $\hat{n}$ as $\hat{n}=\left(\sin a\cos b,\sin a\sin b,\cos a\right)$ using the spherical polar coordinate with $a$ and $b$ ($a,b\in\mathbb{R}$, $a\in[0,\pi]$, $b\in[0,2\pi]$) being respectively the azimuthal and the polar angles, such that  $M_A(k,\hat{n})$ can be represented as $M_A(k,a,b)$. Let us now consider a bipartite state $\rho_{AB}$ of which the partition $A$ is constituted of a single qubit, and the part $B$ can host, in general, multiple qubits. A projection measurement corresponding to $M_A(k,a,b)$ leads to a post-measured ensemble $\mathcal{E}\rightarrow\{p_{k},\varrho_{AB}^{(k)}\}$ on $AB$, where $\varrho_{B}^{(k)}=\text{Tr}_A\left[\varrho_{AB}^{(k)}\right]$, with 
\begin{eqnarray}
    \varrho_{AB}^{(k)}=\frac{1}{p_{(k)}}\left[I_B\otimes P_A^{(k)}\right] \rho_{AB}\left[I_B\otimes P_A^{(k)}\right]
\end{eqnarray}
being the post-measured state on $AB$ corresponding to the measurement outcome $k$, occurring with the probability 
\begin{eqnarray}
    p_{(k)} &=& \text{Tr}\left[\left(I_B\otimes P_A^{(k)}\right) \rho_{AB}\left(I_B\otimes P_A^{(k)}\right)\right]. 
\end{eqnarray}
Here, $P_A^{(k)}$ is defined as $P_A^{(k)}=\ket{\psi_A^{(k)}}\bra{\psi_A^{(k)}}$, $k=\pm1$, where
\begin{eqnarray}
    \ket{\psi_A^{(+1)}} &=& \cos\frac{a}{2}\ket{0}+\text{e}^{\text{i}b}\sin\frac{a}{2}\ket{1},\nonumber\\
    \ket{\psi_A^{(-1)}} &=& \sin\frac{a}{2}\ket{0}-\text{e}^{\text{i}b}\cos\frac{a}{2}\ket{1},
    \label{eq:measurement-basis}
\end{eqnarray}
such that 
\begin{eqnarray}
M_A(k,a,b)&=&P_A^{(k)\dagger} P_A^{(k)},\;
\sum_{k=\pm  1}M_A(k,a,b)=I,
\end{eqnarray} 
with $\{\ket{0},\ket{1}\}$ being the computational basis for the qubit.



\begin{table}
    \centering
    \begin{tabular}{|c|c|c|c|c|c|c|c|c|c|c|c|c|c|c|}
    \hline
    $N$ &\makecell{Target \\ logical \\ state} &$J_2$& $\gamma$ & $a_1$ & $b_1$ & $k_1$ &$a_2$& $b_2$ & $k_2$ & \makecell{$M_{\text{min}}$\\$(\tilde{f}=0.66)$} & \makecell{$p_{(k_1,k_2)}$\\$(\tilde{f}=0.66)$} & \makecell{$M_{\text{min}}$\\$(\tilde{f}=0.9)$} & \makecell{$p_{(k_1,k_2)}$\\$(\tilde{f}=0.9)$}& \makecell{Maximum \\ fidelity \\ achieved} \\ \hline 
        $2$ & $\ket{\Psi_z^+}$ & $0$& $0$ & $0$& - & $+1$& - & - & - & $4$ & $0.36$ & $6$ & $0.287$ & $1$\\ \hline
        $2$ & $\ket{\Psi_z^-}$ & $1$ & $1$ & $0$ & - & $+1$ & - & - & - & $2$ & $0.36$ & $4$ & $0.287$ & $1$ \\ \hline
        $2$ & $\ket{\Psi_x^+}$ & $0$ & $0.85$  & $0.907\pi$ & $0.78\pi$ & $-1$ & - & - & - & 6 & $0.022$ & $12$ & $0.001$ & $0.996$\\ \hline
        $2$ & $\ket{\Psi_x^-}$ & $0$ & $0.85$ & $0.093\pi$ & $0.78\pi$ & $+1$ & - & - & - & 6 & 0.022 & 12 & 0.001 & $0.996$\\ \hline      
        $2$ & $\ket{\Psi_y^+}$ & $0$ & $-0.85$ & $0.093\pi$  & $0.22\pi$ & $+1$ & - & - & - & $7$ & $0.011$ & $13$ & $10^{-4}$ & $0.995$  \\ \hline
        $2$ & $\ket{\Psi_y^-}$ & $0$ & $-0.85$ & $0.0907\pi$ & $0.22\pi$ & $-1$ & - & - & - &  $7$ & $0.011$ & $13$ & $10^{-4}$ & $0.995$\\ \hline 
        $4$ &$\ket{\Psi_z^+}$ & $0$& $0$ & $0$& - & $+1$ & $0$ & - &+1 & $6$ & $0.111$ & $10$ & $0.091$ & $1$ \\ \hline
        $4$ &$\ket{\Psi_z^-}$ & $1$ & $0.17$ & $0$ & - & $+1$& $0$ & - &+1 & $12$ & $0.091$ & $35$ & $0.036$ & $0.993$\\ \hline
        $6$ &$\ket{\Psi_z^+}$ & $0$ & $0$ & $0$ & - & $+1$ & $0$ & -& $+1$ & $15$ & $0.034$ & $24$ & $0.026$ & $1$\\ \hline
        $6$ &$\ket{\Psi_z^-}$ & $1$ & $0.19$ & $0$ & - & $+1$ & $0$ & - &+1 & $24$ & $0.008$ & $47$ & $0.001$ & $0.985$\\ \hline
        $8$ &$\ket{\Psi_z^+}$ & $0$ & $0$ & $0$ & - & $+1$ & $0$ & - &+1 & $31$ & $0.010$ & $50$ & $0.007$ & $1$ \\ \hline
        $8$ &$\ket{\Psi_z^-}$ & $1$ & $0.2$ & $0$ & -& $+1$& $0$ & -&+1 & $80$ & $10^{-6}$ & $223$ & $10^{-11}$ & $0.965$ \\ \hline
        \end{tabular}
    \caption{Details of the purification of logical states in the isotropic Heisenberg model using the EMR protocol with $1$ (for $N=2$) and $2$ (for $N=4,6,8$) auxiliary qubits. All quantities are dimensionless except $(a_j,b_j)$, $j=1,2$, which are in radian.}
    \label{tab:example_table}
\end{table}

\begin{figure*}
    \centering
    \includegraphics[width=\linewidth]{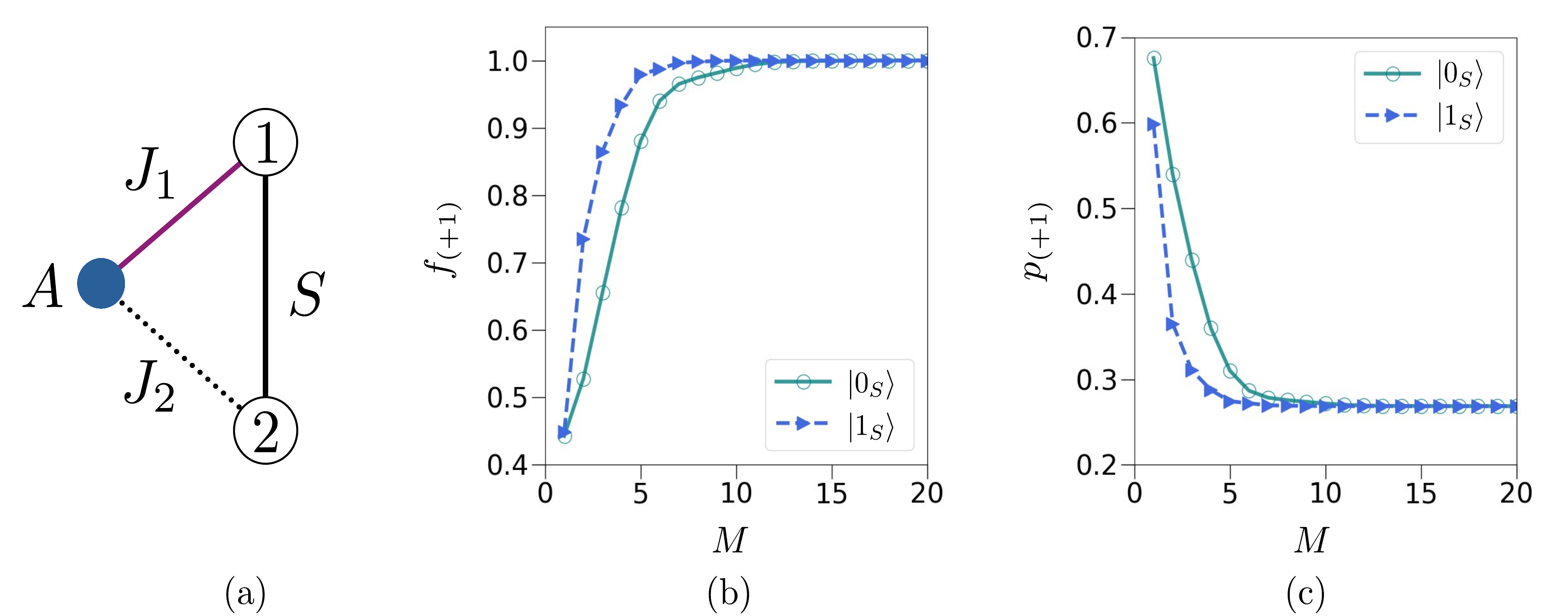}
    \caption{(a) Schematic representation of the interaction strengths between the AQ and the DQs constructing the logical qubit, as represented by the Hamiltonian~(\ref{eq:xy-interaction}), in the case of a two-qubit isotropic Heisenberg model hosting a logical qubit. The variations of (a) $f_{(+1)}$ and $p_{(+1)}$ as functions of the number of rounds, $M$, for the logical states $\ket{0_S}$ and $\ket{1_S}$ (see Eq.~(\ref{eq:heisenberg_logical_state})). All quantities plotted are dimensionless.} 
    \label{fig:minimal_system}
\end{figure*}

\section{Preparation of Heisenberg logical qubit using XY interaction}
\label{app:xy_interaction}

Here, we discuss the preparation of logical states on the $1D$ isotropic Heisenberg Hamiltonian of $N$ ($N$ is even) DQs (see Eq.~(\ref{eq:N_qubit_system_hamiltonian})) using the EMR protocol discussed in Sec.~\ref{sec:repeated_measurements}. For this, we consider $N_A(\leq N)$ auxiliary qubits, with Hamiltonian $H_A=-(E_A/2)\sum_{j=1}^{N_A}\sigma^z_j$, connected to the system such that (see Fig.~\ref{fig:N_qubit_xy})
\begin{enumerate} 
\item[(a)] an auxiliary qubit $a$ is connected to at most one nearest-neighbor (NN) pair $(s-1,s)$ of the system $S$ via the Hamiltonian  
\begin{eqnarray}
    H_{(s,j)}&=&\frac{1}{4}\big[J_1\left(\gamma_+\sigma^x_{j}\sigma^x_{s}+\gamma_-\sigma^y_{j}\sigma^y_{s}\right)+J_2\left(\gamma_+\sigma^x_{j}\sigma^x_{s-1}+\gamma_-\sigma^y_{j}\sigma^y_{s-1}\right)\big],
    \label{eq:unit_Hamiltonian}
\end{eqnarray}
with $J_1$ and $J_2$ being the strengths of the exchange interactions between the qubits $(s,j)$ and $(s-1,j)$ respectively, and $\gamma$ being the $xy$-anisotropy with $\gamma_\pm=(1\pm\gamma)/2$, while 
\item[(b)] the $N_A$ auxiliary qubits are connected to $N_A$ \emph{consecutive} NN pairs of $S$, and 
\item[(c)] one auxiliary qubit $j$ is connected to at most one nearest-neighbor pair $(s-1,s)$ of the system $S$, such that the full system-ancilla Hamiltonian is given by
\begin{eqnarray}
    H_{SA}&=&\sum_{s=1}^N\sum_{j=1}^{N_A}H_{(s,j)}.
    \label{eq:full_system_auxiliary_Hamiltonian}
\end{eqnarray}
\end{enumerate} 
We refer to such a construction of $N$-qubit system $S$ and $N_A$ auxiliary qubits by an $(N,N_A)$-\emph{setup}. We point out here that in the case of $N=2$, the PBC is equivalent to the open boundary condition (OBC), and given conditions (a) and (b), the only possible value of $N_A$ is $1$, giving rise to the $(2,1)$-setup discussed in Sec.~\ref{sec:repeated_measurements} (see Fig.~\ref{fig:minimal_system}). The cardinal states of the logical Bloch sphere in the present case are given by $\ket{\Psi_z^+}=\ket{1_S}$,  $\ket{\Psi_z^-}=\ket{0_S}$, $\ket{\Psi_x^\pm}=(\ket{0_S}\pm\ket{1_S})/\sqrt{2}$ and $\ket{\Psi_y^\pm}=(\ket{0_S}\pm\text{i}\ket{1_S})/\sqrt{2}$, such that $Z\ket{\Psi_z^\pm}=(\pm 1)\ket{\Psi_z^\pm}$, $X\ket{\Psi_x^\pm}=(\pm 1)\ket{\Psi_x^\pm}$, and $Y\ket{\Psi_y^\pm}=(\pm 1)\ket{\Psi_y^\pm}$, where the logical operators $X$, $Y$, and $Z$ are given in Sec.~\ref{sec:designed_interaction}. Our numerical investigation suggests that for $N\geq 4$,  the preparation of logical qubit states using the interaction Hamiltonian~(\ref{eq:full_system_auxiliary_Hamiltonian}) and $N_A=1$ is not possible. Therefore, for $N\geq 4$, we introduce a second auxiliary qubit ($N_A=2$) to facilitate the purification process. The  numerical results corresponding to preparation of different logical states in the case of Heisenberg logical qubit of different sizes (i.e., different $N$) is presented in Table~\ref{tab:example_table}. In the case of $N\geq 4$, each round of purification in the EMR protocol is denoted by $\{t,(a_1,b_1),(a_2,b_2),(k_1,k_2)\}$, where $(a_j,b_j)$ are the real parameters fixing the measurements on the auxiliary qubit $j$, and $k_j$ are the chosen outcome on $j$, with $j=1,2$. For all cases corresponding to $(N,2)$-setup where we are able to purify a logical state, we have considered the purification round to be $\{1,(0,0),(0,0),(+1,+1)\}$, while for the $(2,1)$-setup, it is $\{1,(a,b),k\}$, the specific values of $a,b$, and $k$ given in Table~\ref{tab:example_table}. Note from Table~\ref{tab:example_table} that logical states are \emph{almost perfectly purified} using the EMR protocol when $N$ is low, while obtaining a high fidelity becomes increasingly difficult as $N$ increases, with $N_A$ fixed at $2$. We also point out here that in the case of the $(2,1)$-setup, it is possible to span the entire northern hemisphere of the logical Bloch sphere (i.e., it is possible to prepare a state of the form $\ket{\Psi_S}$ with $0\leq\theta\leq \pi/2$) keeping $J_2=0$ and varying $\gamma$, while spanning the southern hemisphere proves difficult.

\twocolumngrid 

\bibliography{ref}

\begin{thebibliography}{92}%
\makeatletter
\providecommand \@ifxundefined [1]{%
 \@ifx{#1\undefined}
}%
\providecommand \@ifnum [1]{%
 \ifnum #1\expandafter \@firstoftwo
 \else \expandafter \@secondoftwo
 \fi
}%
\providecommand \@ifx [1]{%
 \ifx #1\expandafter \@firstoftwo
 \else \expandafter \@secondoftwo
 \fi
}%
\providecommand \natexlab [1]{#1}%
\providecommand \enquote  [1]{``#1''}%
\providecommand \bibnamefont  [1]{#1}%
\providecommand \bibfnamefont [1]{#1}%
\providecommand \citenamefont [1]{#1}%
\providecommand \href@noop [0]{\@secondoftwo}%
\providecommand \href [0]{\begingroup \@sanitize@url \@href}%
\providecommand \@href[1]{\@@startlink{#1}\@@href}%
\providecommand \@@href[1]{\endgroup#1\@@endlink}%
\providecommand \@sanitize@url [0]{\catcode `\\12\catcode `\$12\catcode
  `\&12\catcode `\#12\catcode `\^12\catcode `\_12\catcode `\%12\relax}%
\providecommand \@@startlink[1]{}%
\providecommand \@@endlink[0]{}%
\providecommand \url  [0]{\begingroup\@sanitize@url \@url }%
\providecommand \@url [1]{\endgroup\@href {#1}{\urlprefix }}%
\providecommand \urlprefix  [0]{URL }%
\providecommand \Eprint [0]{\href }%
\providecommand \doibase [0]{http://dx.doi.org/}%
\providecommand \selectlanguage [0]{\@gobble}%
\providecommand \bibinfo  [0]{\@secondoftwo}%
\providecommand \bibfield  [0]{\@secondoftwo}%
\providecommand \translation [1]{[#1]}%
\providecommand \BibitemOpen [0]{}%
\providecommand \bibitemStop [0]{}%
\providecommand \bibitemNoStop [0]{.\EOS\space}%
\providecommand \EOS [0]{\spacefactor3000\relax}%
\providecommand \BibitemShut  [1]{\csname bibitem#1\endcsname}%
\let\auto@bib@innerbib\@empty
\bibitem [{\citenamefont {Osada}\ \emph {et~al.}(2022)\citenamefont {Osada},
  \citenamefont {Yamazaki},\ and\ \citenamefont {Noguchi}}]{Osada2022}%
  \BibitemOpen
  \bibfield  {author} {\bibinfo {author} {\bibfnamefont {A.}~\bibnamefont
  {Osada}}, \bibinfo {author} {\bibfnamefont {R.}~\bibnamefont {Yamazaki}}, \
  and\ \bibinfo {author} {\bibfnamefont {A.}~\bibnamefont {Noguchi}},\
  }\href@noop {} {\emph {\bibinfo {title} {Introduction to Quantum
  Technologies}}}\ (\bibinfo  {publisher} {Springer Singapore},\ \bibinfo
  {year} {2022})\BibitemShut {NoStop}%
\bibitem [{\citenamefont {Nielsen}\ and\ \citenamefont
  {Chuang}(2010)}]{nielsen2010}%
  \BibitemOpen
  \bibfield  {author} {\bibinfo {author} {\bibfnamefont {M.~A.}\ \bibnamefont
  {Nielsen}}\ and\ \bibinfo {author} {\bibfnamefont {I.~L.}\ \bibnamefont
  {Chuang}},\ }\href@noop {} {\emph {\bibinfo {title} {Quantum Computation and
  Quantum Information}}}\ (\bibinfo  {publisher} {Cambridge University Press},\
  \bibinfo {year} {2010})\BibitemShut {NoStop}%
\bibitem [{\citenamefont {Hayashi}\ \emph {et~al.}(2012)\citenamefont
  {Hayashi}, \citenamefont {Ishizaka}, \citenamefont {Kawachi}, \citenamefont
  {Kimura},\ and\ \citenamefont {Ogawa}}]{Hayashi2012}%
  \BibitemOpen
  \bibfield  {author} {\bibinfo {author} {\bibfnamefont {M.}~\bibnamefont
  {Hayashi}}, \bibinfo {author} {\bibfnamefont {S.}~\bibnamefont {Ishizaka}},
  \bibinfo {author} {\bibfnamefont {A.}~\bibnamefont {Kawachi}}, \bibinfo
  {author} {\bibfnamefont {G.}~\bibnamefont {Kimura}}, \ and\ \bibinfo {author}
  {\bibfnamefont {T.}~\bibnamefont {Ogawa}},\ }\href@noop {} {\emph {\bibinfo
  {title} {Introduction to Quantum Information Science}}}\ (\bibinfo
  {publisher} {Springer Berlin, Heidelberg},\ \bibinfo {year}
  {2012})\BibitemShut {NoStop}%
\bibitem [{\citenamefont {Vedral}(2013)}]{Vedral2013}%
  \BibitemOpen
  \bibfield  {author} {\bibinfo {author} {\bibfnamefont {V.}~\bibnamefont
  {Vedral}},\ }\href@noop {} {\emph {\bibinfo {title} {Introduction to Quantum
  Information Science}}}\ (\bibinfo  {publisher} {Oxford University Press},\
  \bibinfo {year} {2013})\BibitemShut {NoStop}%
\bibitem [{\citenamefont {Wilde}(2013)}]{Wilde_2013}%
  \BibitemOpen
  \bibfield  {author} {\bibinfo {author} {\bibfnamefont {M.~M.}\ \bibnamefont
  {Wilde}},\ }\href@noop {} {\emph {\bibinfo {title} {Quantum Information
  Theory}}}\ (\bibinfo  {publisher} {Cambridge University Press},\ \bibinfo
  {year} {2013})\BibitemShut {NoStop}%
\bibitem [{\citenamefont {Mermin}(2012)}]{Mermin2012}%
  \BibitemOpen
  \bibfield  {author} {\bibinfo {author} {\bibfnamefont {N.~D.}\ \bibnamefont
  {Mermin}},\ }\href@noop {} {\emph {\bibinfo {title} {Quantum Computer
  Science: An Introduction}}}\ (\bibinfo  {publisher} {Cambridge University
  Press},\ \bibinfo {year} {2012})\BibitemShut {NoStop}%
\bibitem [{\citenamefont {Scherer}(2019)}]{Scherer2019}%
  \BibitemOpen
  \bibfield  {author} {\bibinfo {author} {\bibfnamefont {W.}~\bibnamefont
  {Scherer}},\ }\href@noop {} {\emph {\bibinfo {title} {Mathematics of Quantum
  Computing: An Introduction}}}\ (\bibinfo  {publisher} {Springer Cham},\
  \bibinfo {year} {2019})\BibitemShut {NoStop}%
\bibitem [{\citenamefont {Ekert}(1991)}]{Ekert91}%
  \BibitemOpen
  \bibfield  {author} {\bibinfo {author} {\bibfnamefont {A.~K.}\ \bibnamefont
  {Ekert}},\ }\href {\doibase 10.1103/PhysRevLett.67.661} {\bibfield  {journal}
  {\bibinfo  {journal} {Phys. Rev. Lett.}\ }\textbf {\bibinfo {volume} {67}},\
  \bibinfo {pages} {661} (\bibinfo {year} {1991})}\BibitemShut {NoStop}%
\bibitem [{\citenamefont {Bennett}\ \emph {et~al.}(1992)\citenamefont
  {Bennett}, \citenamefont {Brassard},\ and\ \citenamefont
  {Mermin}}]{Bennett1992Crypto}%
  \BibitemOpen
  \bibfield  {author} {\bibinfo {author} {\bibfnamefont {C.~H.}\ \bibnamefont
  {Bennett}}, \bibinfo {author} {\bibfnamefont {G.}~\bibnamefont {Brassard}}, \
  and\ \bibinfo {author} {\bibfnamefont {N.~D.}\ \bibnamefont {Mermin}},\
  }\href {\doibase 10.1103/PhysRevLett.68.557} {\bibfield  {journal} {\bibinfo
  {journal} {Phys. Rev. Lett.}\ }\textbf {\bibinfo {volume} {68}},\ \bibinfo
  {pages} {557} (\bibinfo {year} {1992})}\BibitemShut {NoStop}%
\bibitem [{\citenamefont {Bennett}\ and\ \citenamefont
  {Wiesner}(1992)}]{bennett1992}%
  \BibitemOpen
  \bibfield  {author} {\bibinfo {author} {\bibfnamefont {C.~H.}\ \bibnamefont
  {Bennett}}\ and\ \bibinfo {author} {\bibfnamefont {S.~J.}\ \bibnamefont
  {Wiesner}},\ }\href {\doibase 10.1103/PhysRevLett.69.2881} {\bibfield
  {journal} {\bibinfo  {journal} {Phys. Rev. Lett.}\ }\textbf {\bibinfo
  {volume} {69}},\ \bibinfo {pages} {2881} (\bibinfo {year}
  {1992})}\BibitemShut {NoStop}%
\bibitem [{\citenamefont {Bennett}\ \emph {et~al.}(1993)\citenamefont
  {Bennett}, \citenamefont {Brassard}, \citenamefont {Cr\'epeau}, \citenamefont
  {Jozsa}, \citenamefont {Peres},\ and\ \citenamefont
  {Wootters}}]{bennett1993}%
  \BibitemOpen
  \bibfield  {author} {\bibinfo {author} {\bibfnamefont {C.~H.}\ \bibnamefont
  {Bennett}}, \bibinfo {author} {\bibfnamefont {G.}~\bibnamefont {Brassard}},
  \bibinfo {author} {\bibfnamefont {C.}~\bibnamefont {Cr\'epeau}}, \bibinfo
  {author} {\bibfnamefont {R.}~\bibnamefont {Jozsa}}, \bibinfo {author}
  {\bibfnamefont {A.}~\bibnamefont {Peres}}, \ and\ \bibinfo {author}
  {\bibfnamefont {W.~K.}\ \bibnamefont {Wootters}},\ }\href {\doibase
  10.1103/PhysRevLett.70.1895} {\bibfield  {journal} {\bibinfo  {journal}
  {Phys. Rev. Lett.}\ }\textbf {\bibinfo {volume} {70}},\ \bibinfo {pages}
  {1895} (\bibinfo {year} {1993})}\BibitemShut {NoStop}%
\bibitem [{\citenamefont {Gisin}\ \emph {et~al.}(2002)\citenamefont {Gisin},
  \citenamefont {Ribordy}, \citenamefont {Tittel},\ and\ \citenamefont
  {Zbinden}}]{quantum_crypto_gisin_rmp_2002}%
  \BibitemOpen
  \bibfield  {author} {\bibinfo {author} {\bibfnamefont {N.}~\bibnamefont
  {Gisin}}, \bibinfo {author} {\bibfnamefont {G.}~\bibnamefont {Ribordy}},
  \bibinfo {author} {\bibfnamefont {W.}~\bibnamefont {Tittel}}, \ and\ \bibinfo
  {author} {\bibfnamefont {H.}~\bibnamefont {Zbinden}},\ }\href {\doibase
  10.1103/RevModPhys.74.145} {\bibfield  {journal} {\bibinfo  {journal} {Rev.
  Mod. Phys.}\ }\textbf {\bibinfo {volume} {74}},\ \bibinfo {pages} {145}
  (\bibinfo {year} {2002})}\BibitemShut {NoStop}%
\bibitem [{\citenamefont {Shor}(1997)}]{Shor1997}%
  \BibitemOpen
  \bibfield  {author} {\bibinfo {author} {\bibfnamefont {P.~W.}\ \bibnamefont
  {Shor}},\ }\href {\doibase 10.1137/s0097539795293172} {\bibfield  {journal}
  {\bibinfo  {journal} {{SIAM} Journal on Computing}\ }\textbf {\bibinfo
  {volume} {26}},\ \bibinfo {pages} {1484} (\bibinfo {year}
  {1997})}\BibitemShut {NoStop}%
\bibitem [{\citenamefont {Harrow}\ and\ \citenamefont
  {Montanaro}(2017)}]{Harrow2017}%
  \BibitemOpen
  \bibfield  {author} {\bibinfo {author} {\bibfnamefont {A.~W.}\ \bibnamefont
  {Harrow}}\ and\ \bibinfo {author} {\bibfnamefont {A.}~\bibnamefont
  {Montanaro}},\ }\href {\doibase 10.1038/nature23458} {\bibfield  {journal}
  {\bibinfo  {journal} {Nature}\ }\textbf {\bibinfo {volume} {549}},\ \bibinfo
  {pages} {203} (\bibinfo {year} {2017})}\BibitemShut {NoStop}%
\bibitem [{\citenamefont {Dalzell}\ \emph {et~al.}(2020)\citenamefont
  {Dalzell}, \citenamefont {Harrow}, \citenamefont {Koh},\ and\ \citenamefont
  {La~Placa}}]{Dalzell2020}%
  \BibitemOpen
  \bibfield  {author} {\bibinfo {author} {\bibfnamefont {A.~M.}\ \bibnamefont
  {Dalzell}}, \bibinfo {author} {\bibfnamefont {A.~W.}\ \bibnamefont {Harrow}},
  \bibinfo {author} {\bibfnamefont {D.~E.}\ \bibnamefont {Koh}}, \ and\
  \bibinfo {author} {\bibfnamefont {R.~L.}\ \bibnamefont {La~Placa}},\ }\href
  {\doibase 10.22331/q-2020-05-11-264} {\bibfield  {journal} {\bibinfo
  {journal} {{Quantum}}\ }\textbf {\bibinfo {volume} {4}},\ \bibinfo {pages}
  {264} (\bibinfo {year} {2020})}\BibitemShut {NoStop}%
\bibitem [{\citenamefont {Raussendorf}\ and\ \citenamefont
  {Briegel}(2001)}]{raussendorf2001}%
  \BibitemOpen
  \bibfield  {author} {\bibinfo {author} {\bibfnamefont {R.}~\bibnamefont
  {Raussendorf}}\ and\ \bibinfo {author} {\bibfnamefont {H.~J.}\ \bibnamefont
  {Briegel}},\ }\href {\doibase 10.1103/PhysRevLett.86.5188} {\bibfield
  {journal} {\bibinfo  {journal} {Phys. Rev. Lett.}\ }\textbf {\bibinfo
  {volume} {86}},\ \bibinfo {pages} {5188} (\bibinfo {year}
  {2001})}\BibitemShut {NoStop}%
\bibitem [{\citenamefont {Raussendorf}\ \emph {et~al.}(2003)\citenamefont
  {Raussendorf}, \citenamefont {Browne},\ and\ \citenamefont
  {Briegel}}]{raussendorf2003}%
  \BibitemOpen
  \bibfield  {author} {\bibinfo {author} {\bibfnamefont {R.}~\bibnamefont
  {Raussendorf}}, \bibinfo {author} {\bibfnamefont {D.~E.}\ \bibnamefont
  {Browne}}, \ and\ \bibinfo {author} {\bibfnamefont {H.~J.}\ \bibnamefont
  {Briegel}},\ }\href {\doibase 10.1103/PhysRevA.68.022312} {\bibfield
  {journal} {\bibinfo  {journal} {Phys. Rev. A}\ }\textbf {\bibinfo {volume}
  {68}},\ \bibinfo {pages} {022312} (\bibinfo {year} {2003})}\BibitemShut
  {NoStop}%
\bibitem [{\citenamefont {Mattle}\ \emph {et~al.}(1996)\citenamefont {Mattle},
  \citenamefont {Weinfurter}, \citenamefont {Kwiat},\ and\ \citenamefont
  {Zeilinger}}]{Mattle1996}%
  \BibitemOpen
  \bibfield  {author} {\bibinfo {author} {\bibfnamefont {K.}~\bibnamefont
  {Mattle}}, \bibinfo {author} {\bibfnamefont {H.}~\bibnamefont {Weinfurter}},
  \bibinfo {author} {\bibfnamefont {P.~G.}\ \bibnamefont {Kwiat}}, \ and\
  \bibinfo {author} {\bibfnamefont {A.}~\bibnamefont {Zeilinger}},\ }\href
  {\doibase 10.1103/PhysRevLett.76.4656} {\bibfield  {journal} {\bibinfo
  {journal} {Phys. Rev. Lett.}\ }\textbf {\bibinfo {volume} {76}},\ \bibinfo
  {pages} {4656} (\bibinfo {year} {1996})}\BibitemShut {NoStop}%
\bibitem [{\citenamefont {Bouwmeester}\ \emph {et~al.}(1997)\citenamefont
  {Bouwmeester}, \citenamefont {Pan}, \citenamefont {Mattle}, \citenamefont
  {Eibl}, \citenamefont {Weinfurter},\ and\ \citenamefont
  {Zeilinger}}]{Bouwmeester1997}%
  \BibitemOpen
  \bibfield  {author} {\bibinfo {author} {\bibfnamefont {D.}~\bibnamefont
  {Bouwmeester}}, \bibinfo {author} {\bibfnamefont {J.-W.}\ \bibnamefont
  {Pan}}, \bibinfo {author} {\bibfnamefont {K.}~\bibnamefont {Mattle}},
  \bibinfo {author} {\bibfnamefont {M.}~\bibnamefont {Eibl}}, \bibinfo {author}
  {\bibfnamefont {H.}~\bibnamefont {Weinfurter}}, \ and\ \bibinfo {author}
  {\bibfnamefont {A.}~\bibnamefont {Zeilinger}},\ }\href {\doibase
  10.1038/37539} {\bibfield  {journal} {\bibinfo  {journal} {Nature}\ }\textbf
  {\bibinfo {volume} {390}},\ \bibinfo {pages} {575} (\bibinfo {year}
  {1997})}\BibitemShut {NoStop}%
\bibitem [{\citenamefont {Jennewein}\ \emph {et~al.}(2000)\citenamefont
  {Jennewein}, \citenamefont {Simon}, \citenamefont {Weihs}, \citenamefont
  {Weinfurter},\ and\ \citenamefont {Zeilinger}}]{Jennewein2000}%
  \BibitemOpen
  \bibfield  {author} {\bibinfo {author} {\bibfnamefont {T.}~\bibnamefont
  {Jennewein}}, \bibinfo {author} {\bibfnamefont {C.}~\bibnamefont {Simon}},
  \bibinfo {author} {\bibfnamefont {G.}~\bibnamefont {Weihs}}, \bibinfo
  {author} {\bibfnamefont {H.}~\bibnamefont {Weinfurter}}, \ and\ \bibinfo
  {author} {\bibfnamefont {A.}~\bibnamefont {Zeilinger}},\ }\href {\doibase
  10.1103/PhysRevLett.84.4729} {\bibfield  {journal} {\bibinfo  {journal}
  {Phys. Rev. Lett.}\ }\textbf {\bibinfo {volume} {84}},\ \bibinfo {pages}
  {4729} (\bibinfo {year} {2000})}\BibitemShut {NoStop}%
\bibitem [{\citenamefont {Walther}\ \emph {et~al.}(2005)\citenamefont
  {Walther}, \citenamefont {Resch}, \citenamefont {Rudolph}, \citenamefont
  {Schenck}, \citenamefont {Weinfurter}, \citenamefont {Vedral}, \citenamefont
  {Aspelmeyer},\ and\ \citenamefont {Zeilinger}}]{Walther_2005}%
  \BibitemOpen
  \bibfield  {author} {\bibinfo {author} {\bibfnamefont {P.}~\bibnamefont
  {Walther}}, \bibinfo {author} {\bibfnamefont {K.~J.}\ \bibnamefont {Resch}},
  \bibinfo {author} {\bibfnamefont {T.}~\bibnamefont {Rudolph}}, \bibinfo
  {author} {\bibfnamefont {E.}~\bibnamefont {Schenck}}, \bibinfo {author}
  {\bibfnamefont {H.}~\bibnamefont {Weinfurter}}, \bibinfo {author}
  {\bibfnamefont {V.}~\bibnamefont {Vedral}}, \bibinfo {author} {\bibfnamefont
  {M.}~\bibnamefont {Aspelmeyer}}, \ and\ \bibinfo {author} {\bibfnamefont
  {A.}~\bibnamefont {Zeilinger}},\ }\href {\doibase 10.1038/nature03347}
  {\bibfield  {journal} {\bibinfo  {journal} {Nature}\ }\textbf {\bibinfo
  {volume} {434}},\ \bibinfo {pages} {169} (\bibinfo {year}
  {2005})}\BibitemShut {NoStop}%
\bibitem [{\citenamefont {Lanyon}\ \emph {et~al.}(2007)\citenamefont {Lanyon},
  \citenamefont {Weinhold}, \citenamefont {Langford}, \citenamefont {Barbieri},
  \citenamefont {James}, \citenamefont {Gilchrist},\ and\ \citenamefont
  {White}}]{Lanyon2007}%
  \BibitemOpen
  \bibfield  {author} {\bibinfo {author} {\bibfnamefont {B.~P.}\ \bibnamefont
  {Lanyon}}, \bibinfo {author} {\bibfnamefont {T.~J.}\ \bibnamefont
  {Weinhold}}, \bibinfo {author} {\bibfnamefont {N.~K.}\ \bibnamefont
  {Langford}}, \bibinfo {author} {\bibfnamefont {M.}~\bibnamefont {Barbieri}},
  \bibinfo {author} {\bibfnamefont {D.~F.~V.}\ \bibnamefont {James}}, \bibinfo
  {author} {\bibfnamefont {A.}~\bibnamefont {Gilchrist}}, \ and\ \bibinfo
  {author} {\bibfnamefont {A.~G.}\ \bibnamefont {White}},\ }\href {\doibase
  10.1103/PhysRevLett.99.250505} {\bibfield  {journal} {\bibinfo  {journal}
  {Phys. Rev. Lett.}\ }\textbf {\bibinfo {volume} {99}},\ \bibinfo {pages}
  {250505} (\bibinfo {year} {2007})}\BibitemShut {NoStop}%
\bibitem [{\citenamefont {Lu}\ \emph {et~al.}(2007{\natexlab{a}})\citenamefont
  {Lu}, \citenamefont {Zhou}, \citenamefont {G{\"u}hne}, \citenamefont {Gao},
  \citenamefont {Zhang}, \citenamefont {Yuan}, \citenamefont {Goebel},
  \citenamefont {Yang},\ and\ \citenamefont {Pan}}]{Lu_2007}%
  \BibitemOpen
  \bibfield  {author} {\bibinfo {author} {\bibfnamefont {C.-Y.}\ \bibnamefont
  {Lu}}, \bibinfo {author} {\bibfnamefont {X.-Q.}\ \bibnamefont {Zhou}},
  \bibinfo {author} {\bibfnamefont {O.}~\bibnamefont {G{\"u}hne}}, \bibinfo
  {author} {\bibfnamefont {W.-B.}\ \bibnamefont {Gao}}, \bibinfo {author}
  {\bibfnamefont {J.}~\bibnamefont {Zhang}}, \bibinfo {author} {\bibfnamefont
  {Z.-S.}\ \bibnamefont {Yuan}}, \bibinfo {author} {\bibfnamefont
  {A.}~\bibnamefont {Goebel}}, \bibinfo {author} {\bibfnamefont
  {T.}~\bibnamefont {Yang}}, \ and\ \bibinfo {author} {\bibfnamefont {J.-W.}\
  \bibnamefont {Pan}},\ }\href {\doibase 10.1038/nphys507} {\bibfield
  {journal} {\bibinfo  {journal} {Nature Physics}\ }\textbf {\bibinfo {volume}
  {3}},\ \bibinfo {pages} {91} (\bibinfo {year}
  {2007}{\natexlab{a}})}\BibitemShut {NoStop}%
\bibitem [{\citenamefont {Lu}\ \emph {et~al.}(2007{\natexlab{b}})\citenamefont
  {Lu}, \citenamefont {Browne}, \citenamefont {Yang},\ and\ \citenamefont
  {Pan}}]{Lu_2007a}%
  \BibitemOpen
  \bibfield  {author} {\bibinfo {author} {\bibfnamefont {C.-Y.}\ \bibnamefont
  {Lu}}, \bibinfo {author} {\bibfnamefont {D.~E.}\ \bibnamefont {Browne}},
  \bibinfo {author} {\bibfnamefont {T.}~\bibnamefont {Yang}}, \ and\ \bibinfo
  {author} {\bibfnamefont {J.-W.}\ \bibnamefont {Pan}},\ }\href {\doibase
  10.1103/PhysRevLett.99.250504} {\bibfield  {journal} {\bibinfo  {journal}
  {Phys. Rev. Lett.}\ }\textbf {\bibinfo {volume} {99}},\ \bibinfo {pages}
  {250504} (\bibinfo {year} {2007}{\natexlab{b}})}\BibitemShut {NoStop}%
\bibitem [{\citenamefont {{Romero}}\ and\ \citenamefont
  {{Milburn}}(2024)}]{Romero2024}%
  \BibitemOpen
  \bibfield  {author} {\bibinfo {author} {\bibfnamefont {J.}~\bibnamefont
  {{Romero}}}\ and\ \bibinfo {author} {\bibfnamefont {G.}~\bibnamefont
  {{Milburn}}},\ }\href {\doibase 10.48550/arXiv.2404.03367} {\bibfield
  {journal} {\bibinfo  {journal} {arXiv e-prints}\ } (\bibinfo {year} {2024}),\
  10.48550/arXiv.2404.03367}\BibitemShut {NoStop}%
\bibitem [{\citenamefont {H{\"a}ffner}\ \emph {et~al.}(2008)\citenamefont
  {H{\"a}ffner}, \citenamefont {Roos},\ and\ \citenamefont
  {Blatt}}]{haffner2008}%
  \BibitemOpen
  \bibfield  {author} {\bibinfo {author} {\bibfnamefont {H.}~\bibnamefont
  {H{\"a}ffner}}, \bibinfo {author} {\bibfnamefont {C.}~\bibnamefont {Roos}}, \
  and\ \bibinfo {author} {\bibfnamefont {R.}~\bibnamefont {Blatt}},\ }\href
  {\doibase https://doi.org/10.1016/j.physrep.2008.09.003} {\bibfield
  {journal} {\bibinfo  {journal} {Physics Reports}\ }\textbf {\bibinfo {volume}
  {469}},\ \bibinfo {pages} {155 } (\bibinfo {year} {2008})}\BibitemShut
  {NoStop}%
\bibitem [{\citenamefont {Lanyon}\ \emph {et~al.}(2013)\citenamefont {Lanyon},
  \citenamefont {Jurcevic}, \citenamefont {Zwerger}, \citenamefont {Hempel},
  \citenamefont {Martinez}, \citenamefont {D\"ur}, \citenamefont {Briegel},
  \citenamefont {Blatt},\ and\ \citenamefont {Roos}}]{Lanyon_2013}%
  \BibitemOpen
  \bibfield  {author} {\bibinfo {author} {\bibfnamefont {B.~P.}\ \bibnamefont
  {Lanyon}}, \bibinfo {author} {\bibfnamefont {P.}~\bibnamefont {Jurcevic}},
  \bibinfo {author} {\bibfnamefont {M.}~\bibnamefont {Zwerger}}, \bibinfo
  {author} {\bibfnamefont {C.}~\bibnamefont {Hempel}}, \bibinfo {author}
  {\bibfnamefont {E.~A.}\ \bibnamefont {Martinez}}, \bibinfo {author}
  {\bibfnamefont {W.}~\bibnamefont {D\"ur}}, \bibinfo {author} {\bibfnamefont
  {H.~J.}\ \bibnamefont {Briegel}}, \bibinfo {author} {\bibfnamefont
  {R.}~\bibnamefont {Blatt}}, \ and\ \bibinfo {author} {\bibfnamefont {C.~F.}\
  \bibnamefont {Roos}},\ }\href {\doibase 10.1103/PhysRevLett.111.210501}
  {\bibfield  {journal} {\bibinfo  {journal} {Phys. Rev. Lett.}\ }\textbf
  {\bibinfo {volume} {111}},\ \bibinfo {pages} {210501} (\bibinfo {year}
  {2013})}\BibitemShut {NoStop}%
\bibitem [{\citenamefont {Albarr\'an-Arriagada}\ \emph
  {et~al.}(2018)\citenamefont {Albarr\'an-Arriagada}, \citenamefont
  {Alvarado~Barrios}, \citenamefont {Sanz}, \citenamefont {Romero},
  \citenamefont {Lamata}, \citenamefont {Retamal},\ and\ \citenamefont
  {Solano}}]{Arriagada_2018}%
  \BibitemOpen
  \bibfield  {author} {\bibinfo {author} {\bibfnamefont {F.}~\bibnamefont
  {Albarr\'an-Arriagada}}, \bibinfo {author} {\bibfnamefont {G.}~\bibnamefont
  {Alvarado~Barrios}}, \bibinfo {author} {\bibfnamefont {M.}~\bibnamefont
  {Sanz}}, \bibinfo {author} {\bibfnamefont {G.}~\bibnamefont {Romero}},
  \bibinfo {author} {\bibfnamefont {L.}~\bibnamefont {Lamata}}, \bibinfo
  {author} {\bibfnamefont {J.~C.}\ \bibnamefont {Retamal}}, \ and\ \bibinfo
  {author} {\bibfnamefont {E.}~\bibnamefont {Solano}},\ }\href {\doibase
  10.1103/PhysRevA.97.032320} {\bibfield  {journal} {\bibinfo  {journal} {Phys.
  Rev. A}\ }\textbf {\bibinfo {volume} {97}},\ \bibinfo {pages} {032320}
  (\bibinfo {year} {2018})}\BibitemShut {NoStop}%
\bibitem [{\citenamefont {Krantz}\ \emph {et~al.}(2019)\citenamefont {Krantz},
  \citenamefont {Kjaergaard}, \citenamefont {Yan}, \citenamefont {Orlando},
  \citenamefont {Gustavsson},\ and\ \citenamefont {Oliver}}]{Krantz2019}%
  \BibitemOpen
  \bibfield  {author} {\bibinfo {author} {\bibfnamefont {P.}~\bibnamefont
  {Krantz}}, \bibinfo {author} {\bibfnamefont {M.}~\bibnamefont {Kjaergaard}},
  \bibinfo {author} {\bibfnamefont {F.}~\bibnamefont {Yan}}, \bibinfo {author}
  {\bibfnamefont {T.~P.}\ \bibnamefont {Orlando}}, \bibinfo {author}
  {\bibfnamefont {S.}~\bibnamefont {Gustavsson}}, \ and\ \bibinfo {author}
  {\bibfnamefont {W.~D.}\ \bibnamefont {Oliver}},\ }\href {\doibase
  10.1063/1.5089550} {\bibfield  {journal} {\bibinfo  {journal} {Applied
  Physics Reviews}\ }\textbf {\bibinfo {volume} {6}},\ \bibinfo {pages}
  {021318} (\bibinfo {year} {2019})}\BibitemShut {NoStop}%
\bibitem [{\citenamefont {Kjaergaard}\ \emph {et~al.}(2020)\citenamefont
  {Kjaergaard}, \citenamefont {Schwartz}, \citenamefont {Braum\"uller},
  \citenamefont {Krantz}, \citenamefont {Wang}, \citenamefont {Gustavsson},\
  and\ \citenamefont {Oliver}}]{Kjaergaard2020}%
  \BibitemOpen
  \bibfield  {author} {\bibinfo {author} {\bibfnamefont {M.}~\bibnamefont
  {Kjaergaard}}, \bibinfo {author} {\bibfnamefont {M.~E.}\ \bibnamefont
  {Schwartz}}, \bibinfo {author} {\bibfnamefont {J.}~\bibnamefont
  {Braum\"uller}}, \bibinfo {author} {\bibfnamefont {P.}~\bibnamefont
  {Krantz}}, \bibinfo {author} {\bibfnamefont {J.~I.-J.}\ \bibnamefont {Wang}},
  \bibinfo {author} {\bibfnamefont {S.}~\bibnamefont {Gustavsson}}, \ and\
  \bibinfo {author} {\bibfnamefont {W.~D.}\ \bibnamefont {Oliver}},\ }\href
  {\doibase https://doi.org/10.1146/annurev-conmatphys-031119-050605}
  {\bibfield  {journal} {\bibinfo  {journal} {Annual Review of Condensed Matter
  Physics}\ }\textbf {\bibinfo {volume} {11}},\ \bibinfo {pages} {369}
  (\bibinfo {year} {2020})}\BibitemShut {NoStop}%
\bibitem [{\citenamefont {Bravyi}\ \emph {et~al.}(2022)\citenamefont {Bravyi},
  \citenamefont {Dial}, \citenamefont {Gambetta}, \citenamefont {Gil},\ and\
  \citenamefont {Nazario}}]{Bravyi2022}%
  \BibitemOpen
  \bibfield  {author} {\bibinfo {author} {\bibfnamefont {S.}~\bibnamefont
  {Bravyi}}, \bibinfo {author} {\bibfnamefont {O.}~\bibnamefont {Dial}},
  \bibinfo {author} {\bibfnamefont {J.~M.}\ \bibnamefont {Gambetta}}, \bibinfo
  {author} {\bibfnamefont {D.}~\bibnamefont {Gil}}, \ and\ \bibinfo {author}
  {\bibfnamefont {Z.}~\bibnamefont {Nazario}},\ }\href {\doibase
  10.1063/5.0082975} {\bibfield  {journal} {\bibinfo  {journal} {Journal of
  Applied Physics}\ }\textbf {\bibinfo {volume} {132}},\ \bibinfo {pages}
  {160902} (\bibinfo {year} {2022})}\BibitemShut {NoStop}%
\bibitem [{\citenamefont {Vandersypen}\ \emph {et~al.}(2001)\citenamefont
  {Vandersypen}, \citenamefont {Steffen}, \citenamefont {Breyta}, \citenamefont
  {Yannoni}, \citenamefont {Sherwood},\ and\ \citenamefont
  {Chuang}}]{Vandersypen2001}%
  \BibitemOpen
  \bibfield  {author} {\bibinfo {author} {\bibfnamefont {L.~M.~K.}\
  \bibnamefont {Vandersypen}}, \bibinfo {author} {\bibfnamefont
  {M.}~\bibnamefont {Steffen}}, \bibinfo {author} {\bibfnamefont
  {G.}~\bibnamefont {Breyta}}, \bibinfo {author} {\bibfnamefont {C.~S.}\
  \bibnamefont {Yannoni}}, \bibinfo {author} {\bibfnamefont {M.~H.}\
  \bibnamefont {Sherwood}}, \ and\ \bibinfo {author} {\bibfnamefont {I.~L.}\
  \bibnamefont {Chuang}},\ }\href {\doibase 10.1038/414883a} {\bibfield
  {journal} {\bibinfo  {journal} {Nature}\ }\textbf {\bibinfo {volume} {414}},\
  \bibinfo {pages} {883} (\bibinfo {year} {2001})}\BibitemShut {NoStop}%
\bibitem [{\citenamefont {Ladd}\ \emph {et~al.}(2010)\citenamefont {Ladd},
  \citenamefont {Jelezko}, \citenamefont {Laflamme}, \citenamefont {Nakamura},
  \citenamefont {Monroe},\ and\ \citenamefont {O{'}Brien}}]{Ladd2010Mar}%
  \BibitemOpen
  \bibfield  {author} {\bibinfo {author} {\bibfnamefont {T.~D.}\ \bibnamefont
  {Ladd}}, \bibinfo {author} {\bibfnamefont {F.}~\bibnamefont {Jelezko}},
  \bibinfo {author} {\bibfnamefont {R.}~\bibnamefont {Laflamme}}, \bibinfo
  {author} {\bibfnamefont {Y.}~\bibnamefont {Nakamura}}, \bibinfo {author}
  {\bibfnamefont {C.}~\bibnamefont {Monroe}}, \ and\ \bibinfo {author}
  {\bibfnamefont {J.~L.}\ \bibnamefont {O{'}Brien}},\ }\href {\doibase
  10.1038/nature08812} {\bibfield  {journal} {\bibinfo  {journal} {Nature}\
  }\textbf {\bibinfo {volume} {464}},\ \bibinfo {pages} {45} (\bibinfo {year}
  {2010})}\BibitemShut {NoStop}%
\bibitem [{\citenamefont {{Shor}}(1996)}]{Shor1996a}%
  \BibitemOpen
  \bibfield  {author} {\bibinfo {author} {\bibfnamefont {P.~W.}\ \bibnamefont
  {{Shor}}},\ }\href {\doibase 10.48550/arXiv.quant-ph/9605011} {\bibfield
  {journal} {\bibinfo  {journal} {arXiv e-prints}\ } (\bibinfo {year} {1996}),\
  10.48550/arXiv.quant-ph/9605011}\BibitemShut {NoStop}%
\bibitem [{\citenamefont {Preskill}(1998)}]{preskill1998}%
  \BibitemOpen
  \bibfield  {author} {\bibinfo {author} {\bibfnamefont {J.}~\bibnamefont
  {Preskill}},\ }\enquote {\bibinfo {title} {Fault-tolerant quantum
  computation},}\ in\ \href {\doibase 10.1142/3724} {\emph {\bibinfo
  {booktitle} {Introduction to Quantum Computation and Information}}},\
  \bibinfo {editor} {edited by\ \bibinfo {editor} {\bibfnamefont {H.-K.}\
  \bibnamefont {Lo}}, \bibinfo {editor} {\bibfnamefont {T.}~\bibnamefont
  {Spiller}}, \ and\ \bibinfo {editor} {\bibfnamefont {S.}~\bibnamefont
  {Popescu}}}\ (\bibinfo  {publisher} {WORLD SCIENTIFIC},\ \bibinfo {year}
  {1998})\ Chap.~\bibinfo {chapter} {8}, pp.\ \bibinfo {pages}
  {213--269}\BibitemShut {NoStop}%
\bibitem [{\citenamefont {Gottesman}(2010)}]{gottesman2010}%
  \BibitemOpen
  \bibfield  {author} {\bibinfo {author} {\bibfnamefont {D.}~\bibnamefont
  {Gottesman}},\ }in\ \href {https://arxiv.org/abs/0904.2557} {\emph {\bibinfo
  {booktitle} {Quantum information science and its contributions to
  mathematics, Proceedings of Symposia in Applied Mathematics}}},\
  Vol.~\bibinfo {volume} {68}\ (\bibinfo {year} {2010})\ pp.\ \bibinfo {pages}
  {13--58}\BibitemShut {NoStop}%
\bibitem [{\citenamefont {Terhal}(2015)}]{Terhal2015}%
  \BibitemOpen
  \bibfield  {author} {\bibinfo {author} {\bibfnamefont {B.~M.}\ \bibnamefont
  {Terhal}},\ }\href {\doibase 10.1103/RevModPhys.87.307} {\bibfield  {journal}
  {\bibinfo  {journal} {Rev. Mod. Phys.}\ }\textbf {\bibinfo {volume} {87}},\
  \bibinfo {pages} {307} (\bibinfo {year} {2015})}\BibitemShut {NoStop}%
\bibitem [{\citenamefont {Roffe}(2019)}]{Roffe2019}%
  \BibitemOpen
  \bibfield  {author} {\bibinfo {author} {\bibfnamefont {J.}~\bibnamefont
  {Roffe}},\ }\href {\doibase 10.1080/00107514.2019.1667078} {\bibfield
  {journal} {\bibinfo  {journal} {Contemporary Physics}\ }\textbf {\bibinfo
  {volume} {60}},\ \bibinfo {pages} {226} (\bibinfo {year} {2019})}\BibitemShut
  {NoStop}%
\bibitem [{\citenamefont {Cai}\ \emph {et~al.}(2023)\citenamefont {Cai},
  \citenamefont {Babbush}, \citenamefont {Benjamin}, \citenamefont {Endo},
  \citenamefont {Huggins}, \citenamefont {Li}, \citenamefont {McClean},\ and\
  \citenamefont {O'Brien}}]{Cai2023}%
  \BibitemOpen
  \bibfield  {author} {\bibinfo {author} {\bibfnamefont {Z.}~\bibnamefont
  {Cai}}, \bibinfo {author} {\bibfnamefont {R.}~\bibnamefont {Babbush}},
  \bibinfo {author} {\bibfnamefont {S.~C.}\ \bibnamefont {Benjamin}}, \bibinfo
  {author} {\bibfnamefont {S.}~\bibnamefont {Endo}}, \bibinfo {author}
  {\bibfnamefont {W.~J.}\ \bibnamefont {Huggins}}, \bibinfo {author}
  {\bibfnamefont {Y.}~\bibnamefont {Li}}, \bibinfo {author} {\bibfnamefont
  {J.~R.}\ \bibnamefont {McClean}}, \ and\ \bibinfo {author} {\bibfnamefont
  {T.~E.}\ \bibnamefont {O'Brien}},\ }\href {\doibase
  10.1103/RevModPhys.95.045005} {\bibfield  {journal} {\bibinfo  {journal}
  {Rev. Mod. Phys.}\ }\textbf {\bibinfo {volume} {95}},\ \bibinfo {pages}
  {045005} (\bibinfo {year} {2023})}\BibitemShut {NoStop}%
\bibitem [{\citenamefont {Calderbank}\ and\ \citenamefont
  {Shor}(1996)}]{Calderbank1996}%
  \BibitemOpen
  \bibfield  {author} {\bibinfo {author} {\bibfnamefont {A.~R.}\ \bibnamefont
  {Calderbank}}\ and\ \bibinfo {author} {\bibfnamefont {P.~W.}\ \bibnamefont
  {Shor}},\ }\href {\doibase 10.1103/PhysRevA.54.1098} {\bibfield  {journal}
  {\bibinfo  {journal} {Phys. Rev. A}\ }\textbf {\bibinfo {volume} {54}},\
  \bibinfo {pages} {1098} (\bibinfo {year} {1996})}\BibitemShut {NoStop}%
\bibitem [{\citenamefont {Steane}(1996)}]{Steane1996}%
  \BibitemOpen
  \bibfield  {author} {\bibinfo {author} {\bibfnamefont {A.}~\bibnamefont
  {Steane}},\ }\href {\doibase 10.1098/rspa.1996.0136} {\bibfield  {journal}
  {\bibinfo  {journal} {Proceedings of the Royal Society of London. Series A:
  Mathematical, Physical and Engineering Sciences}\ }\textbf {\bibinfo {volume}
  {452}},\ \bibinfo {pages} {2551} (\bibinfo {year} {1996})},\ \Eprint
  {http://arxiv.org/abs/https://royalsocietypublishing.org/doi/pdf/10.1098/rspa.1996.0136}
  {https://royalsocietypublishing.org/doi/pdf/10.1098/rspa.1996.0136}
  \BibitemShut {NoStop}%
\bibitem [{\citenamefont {Fujii}(2015)}]{fujii2015}%
  \BibitemOpen
  \bibfield  {author} {\bibinfo {author} {\bibfnamefont {K.}~\bibnamefont
  {Fujii}},\ }\href {https://link.springer.com/book/10.1007/978-981-287-996-7}
  {\emph {\bibinfo {title} {{Quantum Computation with Topological Codes}}}}\
  (\bibinfo  {publisher} {Springer Nature},\ \bibinfo {address} {Singapore},\
  \bibinfo {year} {2015})\BibitemShut {NoStop}%
\bibitem [{\citenamefont {Prakash}\ and\ \citenamefont
  {Wei}(2015)}]{prakash_pra_2015}%
  \BibitemOpen
  \bibfield  {author} {\bibinfo {author} {\bibfnamefont {A.}~\bibnamefont
  {Prakash}}\ and\ \bibinfo {author} {\bibfnamefont {T.-C.}\ \bibnamefont
  {Wei}},\ }\href {\doibase 10.1103/PhysRevA.92.022310} {\bibfield  {journal}
  {\bibinfo  {journal} {Phys. Rev. A}\ }\textbf {\bibinfo {volume} {92}},\
  \bibinfo {pages} {022310} (\bibinfo {year} {2015})}\BibitemShut {NoStop}%
\bibitem [{\citenamefont {Dennis}\ \emph {et~al.}(2002)\citenamefont {Dennis},
  \citenamefont {Kitaev}, \citenamefont {Landahl},\ and\ \citenamefont
  {Preskill}}]{kitaev2001}%
  \BibitemOpen
  \bibfield  {author} {\bibinfo {author} {\bibfnamefont {E.}~\bibnamefont
  {Dennis}}, \bibinfo {author} {\bibfnamefont {A.}~\bibnamefont {Kitaev}},
  \bibinfo {author} {\bibfnamefont {A.}~\bibnamefont {Landahl}}, \ and\
  \bibinfo {author} {\bibfnamefont {J.}~\bibnamefont {Preskill}},\ }\href
  {\doibase 10.1063/1.1499754} {\bibfield  {journal} {\bibinfo  {journal} {J.
  Math. Phys.}\ }\textbf {\bibinfo {volume} {43}},\ \bibinfo {pages} {4452}
  (\bibinfo {year} {2002})}\BibitemShut {NoStop}%
\bibitem [{\citenamefont {Kitaev}(2003)}]{kitaev2003}%
  \BibitemOpen
  \bibfield  {author} {\bibinfo {author} {\bibfnamefont {A.}~\bibnamefont
  {Kitaev}},\ }\href {\doibase 10.1016/S0003-4916(02)00018-0} {\bibfield
  {journal} {\bibinfo  {journal} {Ann. Phys.}\ }\textbf {\bibinfo {volume}
  {303}},\ \bibinfo {pages} {2 } (\bibinfo {year} {2003})}\BibitemShut
  {NoStop}%
\bibitem [{\citenamefont {Kitaev}(2006)}]{kitaev2006}%
  \BibitemOpen
  \bibfield  {author} {\bibinfo {author} {\bibfnamefont {A.}~\bibnamefont
  {Kitaev}},\ }\href {\doibase https://doi.org/10.1016/j.aop.2005.10.005}
  {\bibfield  {journal} {\bibinfo  {journal} {Ann. Phys.}\ }\textbf {\bibinfo
  {volume} {321}},\ \bibinfo {pages} {2 } (\bibinfo {year} {2006})}\BibitemShut
  {NoStop}%
\bibitem [{\citenamefont {Bombin}\ and\ \citenamefont
  {Martin-Delgado}(2006)}]{bombin2006}%
  \BibitemOpen
  \bibfield  {author} {\bibinfo {author} {\bibfnamefont {H.}~\bibnamefont
  {Bombin}}\ and\ \bibinfo {author} {\bibfnamefont {M.~A.}\ \bibnamefont
  {Martin-Delgado}},\ }\href {\doibase 10.1103/PhysRevLett.97.180501}
  {\bibfield  {journal} {\bibinfo  {journal} {Phys. Rev. Lett.}\ }\textbf
  {\bibinfo {volume} {97}},\ \bibinfo {pages} {180501} (\bibinfo {year}
  {2006})}\BibitemShut {NoStop}%
\bibitem [{\citenamefont {Bombin}\ and\ \citenamefont
  {Martin-Delgado}(2007)}]{bombin2007}%
  \BibitemOpen
  \bibfield  {author} {\bibinfo {author} {\bibfnamefont {H.}~\bibnamefont
  {Bombin}}\ and\ \bibinfo {author} {\bibfnamefont {M.~A.}\ \bibnamefont
  {Martin-Delgado}},\ }\href {\doibase 10.1103/PhysRevLett.98.160502}
  {\bibfield  {journal} {\bibinfo  {journal} {Phys. Rev. Lett.}\ }\textbf
  {\bibinfo {volume} {98}},\ \bibinfo {pages} {160502} (\bibinfo {year}
  {2007})}\BibitemShut {NoStop}%
\bibitem [{\citenamefont {Preskill}(2018)}]{Preskill_quantum_2018_nisq}%
  \BibitemOpen
  \bibfield  {author} {\bibinfo {author} {\bibfnamefont {J.}~\bibnamefont
  {Preskill}},\ }\href {\doibase 10.22331/q-2018-08-06-79} {\bibfield
  {journal} {\bibinfo  {journal} {{Quantum}}\ }\textbf {\bibinfo {volume}
  {2}},\ \bibinfo {pages} {79} (\bibinfo {year} {2018})}\BibitemShut {NoStop}%
\bibitem [{\citenamefont {Nash}\ \emph {et~al.}(2020)\citenamefont {Nash},
  \citenamefont {Gheorghiu},\ and\ \citenamefont {Mosca}}]{nash2020}%
  \BibitemOpen
  \bibfield  {author} {\bibinfo {author} {\bibfnamefont {B.}~\bibnamefont
  {Nash}}, \bibinfo {author} {\bibfnamefont {V.}~\bibnamefont {Gheorghiu}}, \
  and\ \bibinfo {author} {\bibfnamefont {M.}~\bibnamefont {Mosca}},\ }\href
  {\doibase 10.1088/2058-9565/ab79b1} {\bibfield  {journal} {\bibinfo
  {journal} {Quantum Sci. Technol.}\ }\textbf {\bibinfo {volume} {5}},\
  \bibinfo {pages} {025010} (\bibinfo {year} {2020})}\BibitemShut {NoStop}%
\bibitem [{\citenamefont {Chiaverini}\ \emph {et~al.}(2004)\citenamefont
  {Chiaverini}, \citenamefont {Leibfried}, \citenamefont {Schaetz},
  \citenamefont {Barrett}, \citenamefont {Blakestad}, \citenamefont {Britton},
  \citenamefont {Itano}, \citenamefont {Jost}, \citenamefont {Knill},
  \citenamefont {Langer}, \citenamefont {Ozeri},\ and\ \citenamefont
  {Wineland}}]{Chiaverini2004}%
  \BibitemOpen
  \bibfield  {author} {\bibinfo {author} {\bibfnamefont {J.}~\bibnamefont
  {Chiaverini}}, \bibinfo {author} {\bibfnamefont {D.}~\bibnamefont
  {Leibfried}}, \bibinfo {author} {\bibfnamefont {T.}~\bibnamefont {Schaetz}},
  \bibinfo {author} {\bibfnamefont {M.~D.}\ \bibnamefont {Barrett}}, \bibinfo
  {author} {\bibfnamefont {R.~B.}\ \bibnamefont {Blakestad}}, \bibinfo {author}
  {\bibfnamefont {J.}~\bibnamefont {Britton}}, \bibinfo {author} {\bibfnamefont
  {W.~M.}\ \bibnamefont {Itano}}, \bibinfo {author} {\bibfnamefont {J.~D.}\
  \bibnamefont {Jost}}, \bibinfo {author} {\bibfnamefont {E.}~\bibnamefont
  {Knill}}, \bibinfo {author} {\bibfnamefont {C.}~\bibnamefont {Langer}},
  \bibinfo {author} {\bibfnamefont {R.}~\bibnamefont {Ozeri}}, \ and\ \bibinfo
  {author} {\bibfnamefont {D.~J.}\ \bibnamefont {Wineland}},\ }\href {\doibase
  10.1038/nature03074} {\bibfield  {journal} {\bibinfo  {journal} {Nature}\
  }\textbf {\bibinfo {volume} {432}},\ \bibinfo {pages} {602} (\bibinfo {year}
  {2004})}\BibitemShut {NoStop}%
\bibitem [{\citenamefont {Schindler}\ \emph {et~al.}(2011)\citenamefont
  {Schindler}, \citenamefont {Barreiro}, \citenamefont {Monz}, \citenamefont
  {Nebendahl}, \citenamefont {Nigg}, \citenamefont {Chwalla}, \citenamefont
  {Hennrich},\ and\ \citenamefont {Blatt}}]{Schindler2011}%
  \BibitemOpen
  \bibfield  {author} {\bibinfo {author} {\bibfnamefont {P.}~\bibnamefont
  {Schindler}}, \bibinfo {author} {\bibfnamefont {J.~T.}\ \bibnamefont
  {Barreiro}}, \bibinfo {author} {\bibfnamefont {T.}~\bibnamefont {Monz}},
  \bibinfo {author} {\bibfnamefont {V.}~\bibnamefont {Nebendahl}}, \bibinfo
  {author} {\bibfnamefont {D.}~\bibnamefont {Nigg}}, \bibinfo {author}
  {\bibfnamefont {M.}~\bibnamefont {Chwalla}}, \bibinfo {author} {\bibfnamefont
  {M.}~\bibnamefont {Hennrich}}, \ and\ \bibinfo {author} {\bibfnamefont
  {R.}~\bibnamefont {Blatt}},\ }\href {\doibase 10.1126/science.1203329}
  {\bibfield  {journal} {\bibinfo  {journal} {Science}\ }\textbf {\bibinfo
  {volume} {332}},\ \bibinfo {pages} {1059} (\bibinfo {year} {2011})},\ \Eprint
  {http://arxiv.org/abs/https://www.science.org/doi/pdf/10.1126/science.1203329}
  {https://www.science.org/doi/pdf/10.1126/science.1203329} \BibitemShut
  {NoStop}%
\bibitem [{\citenamefont {Nigg}\ \emph {et~al.}(2014)\citenamefont {Nigg},
  \citenamefont {M{\"u}ller}, \citenamefont {Martinez}, \citenamefont
  {Schindler}, \citenamefont {Hennrich}, \citenamefont {Monz}, \citenamefont
  {Martin-Delgado},\ and\ \citenamefont {Blatt}}]{Nigg2014}%
  \BibitemOpen
  \bibfield  {author} {\bibinfo {author} {\bibfnamefont {D.}~\bibnamefont
  {Nigg}}, \bibinfo {author} {\bibfnamefont {M.}~\bibnamefont {M{\"u}ller}},
  \bibinfo {author} {\bibfnamefont {E.~A.}\ \bibnamefont {Martinez}}, \bibinfo
  {author} {\bibfnamefont {P.}~\bibnamefont {Schindler}}, \bibinfo {author}
  {\bibfnamefont {M.}~\bibnamefont {Hennrich}}, \bibinfo {author}
  {\bibfnamefont {T.}~\bibnamefont {Monz}}, \bibinfo {author} {\bibfnamefont
  {M.~A.}\ \bibnamefont {Martin-Delgado}}, \ and\ \bibinfo {author}
  {\bibfnamefont {R.}~\bibnamefont {Blatt}},\ }\href {\doibase
  10.1126/science.1253742} {\bibfield  {journal} {\bibinfo  {journal}
  {Science}\ }\textbf {\bibinfo {volume} {345}},\ \bibinfo {pages} {302}
  (\bibinfo {year} {2014})}\BibitemShut {NoStop}%
\bibitem [{\citenamefont {Linke}\ \emph {et~al.}(2017)\citenamefont {Linke},
  \citenamefont {Gutierrez}, \citenamefont {Landsman}, \citenamefont {Figgatt},
  \citenamefont {Debnath}, \citenamefont {Brown},\ and\ \citenamefont
  {Monroe}}]{Linke2017}%
  \BibitemOpen
  \bibfield  {author} {\bibinfo {author} {\bibfnamefont {N.~M.}\ \bibnamefont
  {Linke}}, \bibinfo {author} {\bibfnamefont {M.}~\bibnamefont {Gutierrez}},
  \bibinfo {author} {\bibfnamefont {K.~A.}\ \bibnamefont {Landsman}}, \bibinfo
  {author} {\bibfnamefont {C.}~\bibnamefont {Figgatt}}, \bibinfo {author}
  {\bibfnamefont {S.}~\bibnamefont {Debnath}}, \bibinfo {author} {\bibfnamefont
  {K.~R.}\ \bibnamefont {Brown}}, \ and\ \bibinfo {author} {\bibfnamefont
  {C.}~\bibnamefont {Monroe}},\ }\href {\doibase 10.1126/sciadv.1701074}
  {\bibfield  {journal} {\bibinfo  {journal} {Science Advances}\ }\textbf
  {\bibinfo {volume} {3}},\ \bibinfo {pages} {e1701074} (\bibinfo {year}
  {2017})},\ \Eprint
  {http://arxiv.org/abs/https://www.science.org/doi/pdf/10.1126/sciadv.1701074}
  {https://www.science.org/doi/pdf/10.1126/sciadv.1701074} \BibitemShut
  {NoStop}%
\bibitem [{\citenamefont {Pal}\ \emph {et~al.}(2022)\citenamefont {Pal},
  \citenamefont {Schindler}, \citenamefont {Erhard}, \citenamefont {Rivas},
  \citenamefont {Martin-Delgado}, \citenamefont {Blatt}, \citenamefont {Monz},\
  and\ \citenamefont {M{\"{u}}ller}}]{Pal_quantum_2022_relaxation}%
  \BibitemOpen
  \bibfield  {author} {\bibinfo {author} {\bibfnamefont {A.~K.}\ \bibnamefont
  {Pal}}, \bibinfo {author} {\bibfnamefont {P.}~\bibnamefont {Schindler}},
  \bibinfo {author} {\bibfnamefont {A.}~\bibnamefont {Erhard}}, \bibinfo
  {author} {\bibfnamefont {{\'{A}}.}~\bibnamefont {Rivas}}, \bibinfo {author}
  {\bibfnamefont {M.-A.}\ \bibnamefont {Martin-Delgado}}, \bibinfo {author}
  {\bibfnamefont {R.}~\bibnamefont {Blatt}}, \bibinfo {author} {\bibfnamefont
  {T.}~\bibnamefont {Monz}}, \ and\ \bibinfo {author} {\bibfnamefont
  {M.}~\bibnamefont {M{\"{u}}ller}},\ }\href {\doibase
  10.22331/q-2022-01-24-632} {\bibfield  {journal} {\bibinfo  {journal}
  {{Quantum}}\ }\textbf {\bibinfo {volume} {6}},\ \bibinfo {pages} {632}
  (\bibinfo {year} {2022})}\BibitemShut {NoStop}%
\bibitem [{\citenamefont {Reed}\ \emph {et~al.}(2012)\citenamefont {Reed},
  \citenamefont {DiCarlo}, \citenamefont {Nigg}, \citenamefont {Sun},
  \citenamefont {Frunzio}, \citenamefont {Girvin},\ and\ \citenamefont
  {Schoelkopf}}]{Reed2012}%
  \BibitemOpen
  \bibfield  {author} {\bibinfo {author} {\bibfnamefont {M.~D.}\ \bibnamefont
  {Reed}}, \bibinfo {author} {\bibfnamefont {L.}~\bibnamefont {DiCarlo}},
  \bibinfo {author} {\bibfnamefont {S.~E.}\ \bibnamefont {Nigg}}, \bibinfo
  {author} {\bibfnamefont {L.}~\bibnamefont {Sun}}, \bibinfo {author}
  {\bibfnamefont {L.}~\bibnamefont {Frunzio}}, \bibinfo {author} {\bibfnamefont
  {S.~M.}\ \bibnamefont {Girvin}}, \ and\ \bibinfo {author} {\bibfnamefont
  {R.~J.}\ \bibnamefont {Schoelkopf}},\ }\href {\doibase 10.1038/nature10786}
  {\bibfield  {journal} {\bibinfo  {journal} {Nature}\ }\textbf {\bibinfo
  {volume} {482}},\ \bibinfo {pages} {382} (\bibinfo {year}
  {2012})}\BibitemShut {NoStop}%
\bibitem [{\citenamefont {Kelly}\ \emph {et~al.}(2015)\citenamefont {Kelly},
  \citenamefont {Barends}, \citenamefont {Fowler}, \citenamefont {Megrant},
  \citenamefont {Jeffrey}, \citenamefont {White}, \citenamefont {Sank},
  \citenamefont {Mutus}, \citenamefont {Campbell}, \citenamefont {Chen},
  \citenamefont {Chen}, \citenamefont {Chiaro}, \citenamefont {Dunsworth},
  \citenamefont {Hoi}, \citenamefont {Neill}, \citenamefont {O'Malley},
  \citenamefont {Quintana}, \citenamefont {Roushan}, \citenamefont
  {Vainsencher}, \citenamefont {Wenner}, \citenamefont {Cleland},\ and\
  \citenamefont {Martinis}}]{Kelly2015}%
  \BibitemOpen
  \bibfield  {author} {\bibinfo {author} {\bibfnamefont {J.}~\bibnamefont
  {Kelly}}, \bibinfo {author} {\bibfnamefont {R.}~\bibnamefont {Barends}},
  \bibinfo {author} {\bibfnamefont {A.~G.}\ \bibnamefont {Fowler}}, \bibinfo
  {author} {\bibfnamefont {A.}~\bibnamefont {Megrant}}, \bibinfo {author}
  {\bibfnamefont {E.}~\bibnamefont {Jeffrey}}, \bibinfo {author} {\bibfnamefont
  {T.~C.}\ \bibnamefont {White}}, \bibinfo {author} {\bibfnamefont
  {D.}~\bibnamefont {Sank}}, \bibinfo {author} {\bibfnamefont {J.~Y.}\
  \bibnamefont {Mutus}}, \bibinfo {author} {\bibfnamefont {B.}~\bibnamefont
  {Campbell}}, \bibinfo {author} {\bibfnamefont {Y.}~\bibnamefont {Chen}},
  \bibinfo {author} {\bibfnamefont {Z.}~\bibnamefont {Chen}}, \bibinfo {author}
  {\bibfnamefont {B.}~\bibnamefont {Chiaro}}, \bibinfo {author} {\bibfnamefont
  {A.}~\bibnamefont {Dunsworth}}, \bibinfo {author} {\bibfnamefont {I.~C.}\
  \bibnamefont {Hoi}}, \bibinfo {author} {\bibfnamefont {C.}~\bibnamefont
  {Neill}}, \bibinfo {author} {\bibfnamefont {P.~J.~J.}\ \bibnamefont
  {O'Malley}}, \bibinfo {author} {\bibfnamefont {C.}~\bibnamefont {Quintana}},
  \bibinfo {author} {\bibfnamefont {P.}~\bibnamefont {Roushan}}, \bibinfo
  {author} {\bibfnamefont {A.}~\bibnamefont {Vainsencher}}, \bibinfo {author}
  {\bibfnamefont {J.}~\bibnamefont {Wenner}}, \bibinfo {author} {\bibfnamefont
  {A.~N.}\ \bibnamefont {Cleland}}, \ and\ \bibinfo {author} {\bibfnamefont
  {J.~M.}\ \bibnamefont {Martinis}},\ }\href {\doibase 10.1038/nature14270}
  {\bibfield  {journal} {\bibinfo  {journal} {Nature}\ }\textbf {\bibinfo
  {volume} {519}},\ \bibinfo {pages} {66} (\bibinfo {year} {2015})}\BibitemShut
  {NoStop}%
\bibitem [{\citenamefont {Andersen}\ \emph {et~al.}(2020)\citenamefont
  {Andersen}, \citenamefont {Remm}, \citenamefont {Lazar}, \citenamefont
  {Krinner}, \citenamefont {Lacroix}, \citenamefont {Norris}, \citenamefont
  {Gabureac}, \citenamefont {Eichler},\ and\ \citenamefont
  {Wallraff}}]{Andersen2020}%
  \BibitemOpen
  \bibfield  {author} {\bibinfo {author} {\bibfnamefont {C.~K.}\ \bibnamefont
  {Andersen}}, \bibinfo {author} {\bibfnamefont {A.}~\bibnamefont {Remm}},
  \bibinfo {author} {\bibfnamefont {S.}~\bibnamefont {Lazar}}, \bibinfo
  {author} {\bibfnamefont {S.}~\bibnamefont {Krinner}}, \bibinfo {author}
  {\bibfnamefont {N.}~\bibnamefont {Lacroix}}, \bibinfo {author} {\bibfnamefont
  {G.~J.}\ \bibnamefont {Norris}}, \bibinfo {author} {\bibfnamefont
  {M.}~\bibnamefont {Gabureac}}, \bibinfo {author} {\bibfnamefont
  {C.}~\bibnamefont {Eichler}}, \ and\ \bibinfo {author} {\bibfnamefont
  {A.}~\bibnamefont {Wallraff}},\ }\href {\doibase 10.1038/s41567-020-0920-y}
  {\bibfield  {journal} {\bibinfo  {journal} {Nature Physics}\ }\textbf
  {\bibinfo {volume} {16}},\ \bibinfo {pages} {875} (\bibinfo {year}
  {2020})}\BibitemShut {NoStop}%
\bibitem [{\citenamefont {Zhao}\ \emph {et~al.}(2022)\citenamefont {Zhao},
  \citenamefont {Ye}, \citenamefont {Huang}, \citenamefont {Zhang},
  \citenamefont {Wu}, \citenamefont {Guan}, \citenamefont {Zhu}, \citenamefont
  {Wei}, \citenamefont {He}, \citenamefont {Cao}, \citenamefont {Chen},
  \citenamefont {Chung}, \citenamefont {Deng}, \citenamefont {Fan},
  \citenamefont {Gong}, \citenamefont {Guo}, \citenamefont {Guo}, \citenamefont
  {Han}, \citenamefont {Li}, \citenamefont {Li}, \citenamefont {Li},
  \citenamefont {Liang}, \citenamefont {Lin}, \citenamefont {Qian},
  \citenamefont {Rong}, \citenamefont {Su}, \citenamefont {Sun}, \citenamefont
  {Wang}, \citenamefont {Wu}, \citenamefont {Xu}, \citenamefont {Ying},
  \citenamefont {Yu}, \citenamefont {Zha}, \citenamefont {Zhang}, \citenamefont
  {Huo}, \citenamefont {Lu}, \citenamefont {Peng}, \citenamefont {Zhu},\ and\
  \citenamefont {Pan}}]{Zhao2022}%
  \BibitemOpen
  \bibfield  {author} {\bibinfo {author} {\bibfnamefont {Y.}~\bibnamefont
  {Zhao}}, \bibinfo {author} {\bibfnamefont {Y.}~\bibnamefont {Ye}}, \bibinfo
  {author} {\bibfnamefont {H.-L.}\ \bibnamefont {Huang}}, \bibinfo {author}
  {\bibfnamefont {Y.}~\bibnamefont {Zhang}}, \bibinfo {author} {\bibfnamefont
  {D.}~\bibnamefont {Wu}}, \bibinfo {author} {\bibfnamefont {H.}~\bibnamefont
  {Guan}}, \bibinfo {author} {\bibfnamefont {Q.}~\bibnamefont {Zhu}}, \bibinfo
  {author} {\bibfnamefont {Z.}~\bibnamefont {Wei}}, \bibinfo {author}
  {\bibfnamefont {T.}~\bibnamefont {He}}, \bibinfo {author} {\bibfnamefont
  {S.}~\bibnamefont {Cao}}, \bibinfo {author} {\bibfnamefont {F.}~\bibnamefont
  {Chen}}, \bibinfo {author} {\bibfnamefont {T.-H.}\ \bibnamefont {Chung}},
  \bibinfo {author} {\bibfnamefont {H.}~\bibnamefont {Deng}}, \bibinfo {author}
  {\bibfnamefont {D.}~\bibnamefont {Fan}}, \bibinfo {author} {\bibfnamefont
  {M.}~\bibnamefont {Gong}}, \bibinfo {author} {\bibfnamefont {C.}~\bibnamefont
  {Guo}}, \bibinfo {author} {\bibfnamefont {S.}~\bibnamefont {Guo}}, \bibinfo
  {author} {\bibfnamefont {L.}~\bibnamefont {Han}}, \bibinfo {author}
  {\bibfnamefont {N.}~\bibnamefont {Li}}, \bibinfo {author} {\bibfnamefont
  {S.}~\bibnamefont {Li}}, \bibinfo {author} {\bibfnamefont {Y.}~\bibnamefont
  {Li}}, \bibinfo {author} {\bibfnamefont {F.}~\bibnamefont {Liang}}, \bibinfo
  {author} {\bibfnamefont {J.}~\bibnamefont {Lin}}, \bibinfo {author}
  {\bibfnamefont {H.}~\bibnamefont {Qian}}, \bibinfo {author} {\bibfnamefont
  {H.}~\bibnamefont {Rong}}, \bibinfo {author} {\bibfnamefont {H.}~\bibnamefont
  {Su}}, \bibinfo {author} {\bibfnamefont {L.}~\bibnamefont {Sun}}, \bibinfo
  {author} {\bibfnamefont {S.}~\bibnamefont {Wang}}, \bibinfo {author}
  {\bibfnamefont {Y.}~\bibnamefont {Wu}}, \bibinfo {author} {\bibfnamefont
  {Y.}~\bibnamefont {Xu}}, \bibinfo {author} {\bibfnamefont {C.}~\bibnamefont
  {Ying}}, \bibinfo {author} {\bibfnamefont {J.}~\bibnamefont {Yu}}, \bibinfo
  {author} {\bibfnamefont {C.}~\bibnamefont {Zha}}, \bibinfo {author}
  {\bibfnamefont {K.}~\bibnamefont {Zhang}}, \bibinfo {author} {\bibfnamefont
  {Y.-H.}\ \bibnamefont {Huo}}, \bibinfo {author} {\bibfnamefont {C.-Y.}\
  \bibnamefont {Lu}}, \bibinfo {author} {\bibfnamefont {C.-Z.}\ \bibnamefont
  {Peng}}, \bibinfo {author} {\bibfnamefont {X.}~\bibnamefont {Zhu}}, \ and\
  \bibinfo {author} {\bibfnamefont {J.-W.}\ \bibnamefont {Pan}},\ }\href
  {\doibase 10.1103/PhysRevLett.129.030501} {\bibfield  {journal} {\bibinfo
  {journal} {Phys. Rev. Lett.}\ }\textbf {\bibinfo {volume} {129}},\ \bibinfo
  {pages} {030501} (\bibinfo {year} {2022})}\BibitemShut {NoStop}%
\bibitem [{\citenamefont {Krinner}\ \emph {et~al.}(2022)\citenamefont
  {Krinner}, \citenamefont {Lacroix}, \citenamefont {Remm}, \citenamefont
  {Di~Paolo}, \citenamefont {Genois}, \citenamefont {Leroux}, \citenamefont
  {Hellings}, \citenamefont {Lazar}, \citenamefont {Swiadek}, \citenamefont
  {Herrmann}, \citenamefont {Norris}, \citenamefont {Andersen}, \citenamefont
  {M{\ifmmode\ddot{u}\else\"{u}\fi}ller}, \citenamefont {Blais}, \citenamefont
  {Eichler},\ and\ \citenamefont {Wallraff}}]{Krinner2022May}%
  \BibitemOpen
  \bibfield  {author} {\bibinfo {author} {\bibfnamefont {S.}~\bibnamefont
  {Krinner}}, \bibinfo {author} {\bibfnamefont {N.}~\bibnamefont {Lacroix}},
  \bibinfo {author} {\bibfnamefont {A.}~\bibnamefont {Remm}}, \bibinfo {author}
  {\bibfnamefont {A.}~\bibnamefont {Di~Paolo}}, \bibinfo {author}
  {\bibfnamefont {E.}~\bibnamefont {Genois}}, \bibinfo {author} {\bibfnamefont
  {C.}~\bibnamefont {Leroux}}, \bibinfo {author} {\bibfnamefont
  {C.}~\bibnamefont {Hellings}}, \bibinfo {author} {\bibfnamefont
  {S.}~\bibnamefont {Lazar}}, \bibinfo {author} {\bibfnamefont
  {F.}~\bibnamefont {Swiadek}}, \bibinfo {author} {\bibfnamefont
  {J.}~\bibnamefont {Herrmann}}, \bibinfo {author} {\bibfnamefont {G.~J.}\
  \bibnamefont {Norris}}, \bibinfo {author} {\bibfnamefont {C.~K.}\
  \bibnamefont {Andersen}}, \bibinfo {author} {\bibfnamefont {M.}~\bibnamefont
  {M{\ifmmode\ddot{u}\else\"{u}\fi}ller}}, \bibinfo {author} {\bibfnamefont
  {A.}~\bibnamefont {Blais}}, \bibinfo {author} {\bibfnamefont
  {C.}~\bibnamefont {Eichler}}, \ and\ \bibinfo {author} {\bibfnamefont
  {A.}~\bibnamefont {Wallraff}},\ }\href {\doibase 10.1038/s41586-022-04566-8}
  {\bibfield  {journal} {\bibinfo  {journal} {Nature}\ }\textbf {\bibinfo
  {volume} {605}},\ \bibinfo {pages} {669} (\bibinfo {year}
  {2022})}\BibitemShut {NoStop}%
\bibitem [{\citenamefont {{Acharya}}\ \emph {et~al.}(2024)\citenamefont
  {{Acharya}}, \citenamefont {{Aghababaie-Beni}}, \citenamefont {{Aleiner}}
  \emph {et~al.}}]{Acharya2024}%
  \BibitemOpen
  \bibfield  {author} {\bibinfo {author} {\bibfnamefont {R.}~\bibnamefont
  {{Acharya}}}, \bibinfo {author} {\bibfnamefont {L.}~\bibnamefont
  {{Aghababaie-Beni}}}, \bibinfo {author} {\bibfnamefont {I.}~\bibnamefont
  {{Aleiner}}},  \emph {et~al.},\ }\href {\doibase 10.48550/arXiv.2408.13687}
  {\bibfield  {journal} {\bibinfo  {journal} {arXiv e-prints}\ } (\bibinfo
  {year} {2024}),\ 10.48550/arXiv.2408.13687}\BibitemShut {NoStop}%
\bibitem [{\citenamefont {Gao}\ \emph {et~al.}(2024)\citenamefont {Gao},
  \citenamefont {Fan}, \citenamefont {Zha} \emph {et~al.}}]{gao2024}%
  \BibitemOpen
  \bibfield  {author} {\bibinfo {author} {\bibfnamefont {D.}~\bibnamefont
  {Gao}}, \bibinfo {author} {\bibfnamefont {D.}~\bibnamefont {Fan}}, \bibinfo
  {author} {\bibfnamefont {C.}~\bibnamefont {Zha}},  \emph {et~al.},\ }\href
  {\doibase 10.48550/arXiv.2412.11924} {\bibfield  {journal} {\bibinfo
  {journal} {arXiv}\ } (\bibinfo {year} {2024}),\ 10.48550/arXiv.2412.11924},\
  \Eprint {http://arxiv.org/abs/2412.11924} {2412.11924} \BibitemShut {NoStop}%
\bibitem [{\citenamefont {Bose}(2003)}]{Bose03}%
  \BibitemOpen
  \bibfield  {author} {\bibinfo {author} {\bibfnamefont {S.}~\bibnamefont
  {Bose}},\ }\href {\doibase 10.1103/PhysRevLett.91.207901} {\bibfield
  {journal} {\bibinfo  {journal} {Phys. Rev. Lett.}\ }\textbf {\bibinfo
  {volume} {91}},\ \bibinfo {pages} {207901} (\bibinfo {year}
  {2003})}\BibitemShut {NoStop}%
\bibitem [{\citenamefont {Pushpan}\ \emph
  {et~al.}(2024{\natexlab{a}})\citenamefont {Pushpan}, \citenamefont {K~J},\
  and\ \citenamefont {Pal}}]{Pushpan2024Jul}%
  \BibitemOpen
  \bibfield  {author} {\bibinfo {author} {\bibfnamefont {C.~B.}\ \bibnamefont
  {Pushpan}}, \bibinfo {author} {\bibfnamefont {H.}~\bibnamefont {K~J}}, \ and\
  \bibinfo {author} {\bibfnamefont {A.~K.}\ \bibnamefont {Pal}},\ }\href
  {\doibase 10.1016/j.physleta.2024.129543} {\bibfield  {journal} {\bibinfo
  {journal} {Phys. Lett. A}\ }\textbf {\bibinfo {volume} {511}},\ \bibinfo
  {pages} {129543} (\bibinfo {year} {2024}{\natexlab{a}})}\BibitemShut
  {NoStop}%
\bibitem [{\citenamefont {Fowler}\ \emph {et~al.}(2012)\citenamefont {Fowler},
  \citenamefont {Mariantoni}, \citenamefont {Martinis},\ and\ \citenamefont
  {Cleland}}]{Fowler2012}%
  \BibitemOpen
  \bibfield  {author} {\bibinfo {author} {\bibfnamefont {A.~G.}\ \bibnamefont
  {Fowler}}, \bibinfo {author} {\bibfnamefont {M.}~\bibnamefont {Mariantoni}},
  \bibinfo {author} {\bibfnamefont {J.~M.}\ \bibnamefont {Martinis}}, \ and\
  \bibinfo {author} {\bibfnamefont {A.~N.}\ \bibnamefont {Cleland}},\ }\href
  {\doibase 10.1103/PhysRevA.86.032324} {\bibfield  {journal} {\bibinfo
  {journal} {Phys. Rev. A}\ }\textbf {\bibinfo {volume} {86}},\ \bibinfo
  {pages} {032324} (\bibinfo {year} {2012})}\BibitemShut {NoStop}%
\bibitem [{\citenamefont {Gidney}(2024)}]{Gidney2024}%
  \BibitemOpen
  \bibfield  {author} {\bibinfo {author} {\bibfnamefont {C.}~\bibnamefont
  {Gidney}},\ }\href {\doibase 10.22331/q-2024-04-08-1310} {\bibfield
  {journal} {\bibinfo  {journal} {{Quantum}}\ }\textbf {\bibinfo {volume}
  {8}},\ \bibinfo {pages} {1310} (\bibinfo {year} {2024})}\BibitemShut
  {NoStop}%
\bibitem [{\citenamefont {Aliferis}\ and\ \citenamefont
  {Terhal}(2007)}]{Aliferis2007}%
  \BibitemOpen
  \bibfield  {author} {\bibinfo {author} {\bibfnamefont {P.}~\bibnamefont
  {Aliferis}}\ and\ \bibinfo {author} {\bibfnamefont {B.~M.}\ \bibnamefont
  {Terhal}},\ }\href@noop {} {\bibfield  {journal} {\bibinfo  {journal}
  {Quantum Info. Comput.}\ }\textbf {\bibinfo {volume} {7}},\ \bibinfo {pages}
  {139–156} (\bibinfo {year} {2007})}\BibitemShut {NoStop}%
\bibitem [{\citenamefont {Miao}\ \emph {et~al.}(2023)\citenamefont {Miao},
  \citenamefont {McEwen}, \citenamefont {Atalaya}, \citenamefont {Kafri},
  \citenamefont {Pryadko} \emph {et~al.}}]{Miao2023}%
  \BibitemOpen
  \bibfield  {author} {\bibinfo {author} {\bibfnamefont {K.~C.}\ \bibnamefont
  {Miao}}, \bibinfo {author} {\bibfnamefont {M.}~\bibnamefont {McEwen}},
  \bibinfo {author} {\bibfnamefont {J.}~\bibnamefont {Atalaya}}, \bibinfo
  {author} {\bibfnamefont {D.}~\bibnamefont {Kafri}}, \bibinfo {author}
  {\bibfnamefont {L.~P.}\ \bibnamefont {Pryadko}},  \emph {et~al.},\ }\href
  {\doibase 10.1038/s41567-023-02226-w} {\bibfield  {journal} {\bibinfo
  {journal} {Nature Physics}\ }\textbf {\bibinfo {volume} {19}},\ \bibinfo
  {pages} {1780} (\bibinfo {year} {2023})}\BibitemShut {NoStop}%
\bibitem [{\citenamefont {{Manabe}}\ \emph {et~al.}(2023)\citenamefont
  {{Manabe}}, \citenamefont {{Suzuki}},\ and\ \citenamefont
  {{Darmawan}}}]{Manabe2023}%
  \BibitemOpen
  \bibfield  {author} {\bibinfo {author} {\bibfnamefont {H.}~\bibnamefont
  {{Manabe}}}, \bibinfo {author} {\bibfnamefont {Y.}~\bibnamefont {{Suzuki}}},
  \ and\ \bibinfo {author} {\bibfnamefont {A.~S.}\ \bibnamefont {{Darmawan}}},\
  }\href {\doibase 10.48550/arXiv.2308.08186} {\bibfield  {journal} {\bibinfo
  {journal} {arXiv e-prints}\ ,\ \bibinfo {eid} {arXiv:2308.08186}} (\bibinfo
  {year} {2023})},\ \Eprint {http://arxiv.org/abs/2308.08186} {arXiv:2308.08186
  [quant-ph]} \BibitemShut {NoStop}%
\bibitem [{\citenamefont {Bravyi}\ and\ \citenamefont
  {Kitaev}(2005)}]{Bravyi2005}%
  \BibitemOpen
  \bibfield  {author} {\bibinfo {author} {\bibfnamefont {S.}~\bibnamefont
  {Bravyi}}\ and\ \bibinfo {author} {\bibfnamefont {A.}~\bibnamefont
  {Kitaev}},\ }\href {\doibase 10.1103/PhysRevA.71.022316} {\bibfield
  {journal} {\bibinfo  {journal} {Phys. Rev. A}\ }\textbf {\bibinfo {volume}
  {71}},\ \bibinfo {pages} {022316} (\bibinfo {year} {2005})}\BibitemShut
  {NoStop}%
\bibitem [{\citenamefont {Li}(2015)}]{Li_2015}%
  \BibitemOpen
  \bibfield  {author} {\bibinfo {author} {\bibfnamefont {Y.}~\bibnamefont
  {Li}},\ }\href {\doibase 10.1088/1367-2630/17/2/023037} {\bibfield  {journal}
  {\bibinfo  {journal} {New Journal of Physics}\ }\textbf {\bibinfo {volume}
  {17}},\ \bibinfo {pages} {023037} (\bibinfo {year} {2015})}\BibitemShut
  {NoStop}%
\bibitem [{\citenamefont {Campbell}\ and\ \citenamefont
  {Howard}(2017)}]{Campbell2017}%
  \BibitemOpen
  \bibfield  {author} {\bibinfo {author} {\bibfnamefont {E.~T.}\ \bibnamefont
  {Campbell}}\ and\ \bibinfo {author} {\bibfnamefont {M.}~\bibnamefont
  {Howard}},\ }\href {\doibase 10.1103/PhysRevA.95.022316} {\bibfield
  {journal} {\bibinfo  {journal} {Phys. Rev. A}\ }\textbf {\bibinfo {volume}
  {95}},\ \bibinfo {pages} {022316} (\bibinfo {year} {2017})}\BibitemShut
  {NoStop}%
\bibitem [{\citenamefont {Gidney}\ and\ \citenamefont
  {Fowler}(2019)}]{Gidney2019}%
  \BibitemOpen
  \bibfield  {author} {\bibinfo {author} {\bibfnamefont {C.}~\bibnamefont
  {Gidney}}\ and\ \bibinfo {author} {\bibfnamefont {A.~G.}\ \bibnamefont
  {Fowler}},\ }\href {\doibase 10.22331/q-2019-04-30-135} {\bibfield  {journal}
  {\bibinfo  {journal} {{Quantum}}\ }\textbf {\bibinfo {volume} {3}},\ \bibinfo
  {pages} {135} (\bibinfo {year} {2019})}\BibitemShut {NoStop}%
\bibitem [{\citenamefont {{Sales Rodriguez}}\ \emph {et~al.}(2024)\citenamefont
  {{Sales Rodriguez}}, \citenamefont {{Robinson}},\ and\ \citenamefont
  {{Jepsen}}}]{Rodriguez2024}%
  \BibitemOpen
  \bibfield  {author} {\bibinfo {author} {\bibfnamefont {P.}~\bibnamefont
  {{Sales Rodriguez}}}, \bibinfo {author} {\bibfnamefont {J.~M.}\ \bibnamefont
  {{Robinson}}}, \ and\ \bibinfo {author} {\bibfnamefont {P.~N.~o.}\
  \bibnamefont {{Jepsen}}},\ }\href {\doibase 10.48550/arXiv.2412.15165}
  {\bibfield  {journal} {\bibinfo  {journal} {arXiv e-prints}\ ,\ \bibinfo
  {eid} {arXiv:2412.15165}} (\bibinfo {year} {2024})},\ \Eprint
  {http://arxiv.org/abs/2412.15165} {arXiv:2412.15165 [quant-ph]} \BibitemShut
  {NoStop}%
\bibitem [{\citenamefont {Yan}\ and\ \citenamefont
  {Jing}(2023)}]{jing_measure_ghz}%
  \BibitemOpen
  \bibfield  {author} {\bibinfo {author} {\bibfnamefont {J.-s.}\ \bibnamefont
  {Yan}}\ and\ \bibinfo {author} {\bibfnamefont {J.}~\bibnamefont {Jing}},\
  }\href {\doibase 10.1103/PhysRevA.108.042215} {\bibfield  {journal} {\bibinfo
   {journal} {Phys. Rev. A}\ }\textbf {\bibinfo {volume} {108}},\ \bibinfo
  {pages} {042215} (\bibinfo {year} {2023})}\BibitemShut {NoStop}%
\bibitem [{\citenamefont {Nakazato}\ \emph {et~al.}(2003)\citenamefont
  {Nakazato}, \citenamefont {Takazawa},\ and\ \citenamefont
  {Yuasa}}]{Nakazato_2003}%
  \BibitemOpen
  \bibfield  {author} {\bibinfo {author} {\bibfnamefont {H.}~\bibnamefont
  {Nakazato}}, \bibinfo {author} {\bibfnamefont {T.}~\bibnamefont {Takazawa}},
  \ and\ \bibinfo {author} {\bibfnamefont {K.}~\bibnamefont {Yuasa}},\ }\href
  {\doibase 10.1103/PhysRevLett.90.060401} {\bibfield  {journal} {\bibinfo
  {journal} {Phys. Rev. Lett.}\ }\textbf {\bibinfo {volume} {90}},\ \bibinfo
  {pages} {060401} (\bibinfo {year} {2003})}\BibitemShut {NoStop}%
\bibitem [{\citenamefont {Nakazato}\ \emph {et~al.}(2004)\citenamefont
  {Nakazato}, \citenamefont {Unoki},\ and\ \citenamefont
  {Yuasa}}]{Nakazato_2004}%
  \BibitemOpen
  \bibfield  {author} {\bibinfo {author} {\bibfnamefont {H.}~\bibnamefont
  {Nakazato}}, \bibinfo {author} {\bibfnamefont {M.}~\bibnamefont {Unoki}}, \
  and\ \bibinfo {author} {\bibfnamefont {K.}~\bibnamefont {Yuasa}},\ }\href
  {\doibase 10.1103/PhysRevA.70.012303} {\bibfield  {journal} {\bibinfo
  {journal} {Phys. Rev. A}\ }\textbf {\bibinfo {volume} {70}},\ \bibinfo
  {pages} {012303} (\bibinfo {year} {2004})}\BibitemShut {NoStop}%
\bibitem [{\citenamefont {Roy}\ \emph {et~al.}(2020)\citenamefont {Roy},
  \citenamefont {Chalker}, \citenamefont {Gornyi},\ and\ \citenamefont
  {Gefen}}]{gefen_prr_2020}%
  \BibitemOpen
  \bibfield  {author} {\bibinfo {author} {\bibfnamefont {S.}~\bibnamefont
  {Roy}}, \bibinfo {author} {\bibfnamefont {J.~T.}\ \bibnamefont {Chalker}},
  \bibinfo {author} {\bibfnamefont {I.~V.}\ \bibnamefont {Gornyi}}, \ and\
  \bibinfo {author} {\bibfnamefont {Y.}~\bibnamefont {Gefen}},\ }\href
  {\doibase 10.1103/PhysRevResearch.2.033347} {\bibfield  {journal} {\bibinfo
  {journal} {Phys. Rev. Res.}\ }\textbf {\bibinfo {volume} {2}},\ \bibinfo
  {pages} {033347} (\bibinfo {year} {2020})}\BibitemShut {NoStop}%
\bibitem [{\citenamefont {Langbehn}\ \emph {et~al.}(2024)\citenamefont
  {Langbehn}, \citenamefont {Snizhko}, \citenamefont {Gornyi}, \citenamefont
  {Morigi}, \citenamefont {Gefen},\ and\ \citenamefont
  {Koch}}]{gefen_prx_2024}%
  \BibitemOpen
  \bibfield  {author} {\bibinfo {author} {\bibfnamefont {J.}~\bibnamefont
  {Langbehn}}, \bibinfo {author} {\bibfnamefont {K.}~\bibnamefont {Snizhko}},
  \bibinfo {author} {\bibfnamefont {I.}~\bibnamefont {Gornyi}}, \bibinfo
  {author} {\bibfnamefont {G.}~\bibnamefont {Morigi}}, \bibinfo {author}
  {\bibfnamefont {Y.}~\bibnamefont {Gefen}}, \ and\ \bibinfo {author}
  {\bibfnamefont {C.~P.}\ \bibnamefont {Koch}},\ }\href {\doibase
  10.1103/PRXQuantum.5.030301} {\bibfield  {journal} {\bibinfo  {journal} {PRX
  Quantum}\ }\textbf {\bibinfo {volume} {5}},\ \bibinfo {pages} {030301}
  (\bibinfo {year} {2024})}\BibitemShut {NoStop}%
\bibitem [{\citenamefont {Massar}\ and\ \citenamefont
  {Popescu}(1995)}]{MassarPopescu95}%
  \BibitemOpen
  \bibfield  {author} {\bibinfo {author} {\bibfnamefont {S.}~\bibnamefont
  {Massar}}\ and\ \bibinfo {author} {\bibfnamefont {S.}~\bibnamefont
  {Popescu}},\ }\href {\doibase 10.1103/PhysRevLett.74.1259} {\bibfield
  {journal} {\bibinfo  {journal} {Phys. Rev. Lett.}\ }\textbf {\bibinfo
  {volume} {74}},\ \bibinfo {pages} {1259} (\bibinfo {year}
  {1995})}\BibitemShut {NoStop}%
\bibitem [{\citenamefont {Mintert}\ and\ \citenamefont
  {Wunderlich}(2001)}]{Mintert2001}%
  \BibitemOpen
  \bibfield  {author} {\bibinfo {author} {\bibfnamefont {F.}~\bibnamefont
  {Mintert}}\ and\ \bibinfo {author} {\bibfnamefont {C.}~\bibnamefont
  {Wunderlich}},\ }\href {\doibase 10.1103/PhysRevLett.87.257904} {\bibfield
  {journal} {\bibinfo  {journal} {Phys. Rev. Lett.}\ }\textbf {\bibinfo
  {volume} {87}},\ \bibinfo {pages} {257904} (\bibinfo {year}
  {2001})}\BibitemShut {NoStop}%
\bibitem [{\citenamefont {Porras}\ and\ \citenamefont
  {Cirac}(2004)}]{porras2004}%
  \BibitemOpen
  \bibfield  {author} {\bibinfo {author} {\bibfnamefont {D.}~\bibnamefont
  {Porras}}\ and\ \bibinfo {author} {\bibfnamefont {J.~I.}\ \bibnamefont
  {Cirac}},\ }\href {\doibase 10.1103/PhysRevLett.92.207901} {\bibfield
  {journal} {\bibinfo  {journal} {Phys. Rev. Lett.}\ }\textbf {\bibinfo
  {volume} {92}},\ \bibinfo {pages} {207901} (\bibinfo {year}
  {2004})}\BibitemShut {NoStop}%
\bibitem [{\citenamefont {Dalmonte}\ \emph {et~al.}(2015)\citenamefont
  {Dalmonte}, \citenamefont {Mirzaei}, \citenamefont {Muppalla}, \citenamefont
  {Marcos}, \citenamefont {Zoller},\ and\ \citenamefont
  {Kirchmair}}]{dalmonte2015}%
  \BibitemOpen
  \bibfield  {author} {\bibinfo {author} {\bibfnamefont {M.}~\bibnamefont
  {Dalmonte}}, \bibinfo {author} {\bibfnamefont {S.~I.}\ \bibnamefont
  {Mirzaei}}, \bibinfo {author} {\bibfnamefont {P.~R.}\ \bibnamefont
  {Muppalla}}, \bibinfo {author} {\bibfnamefont {D.}~\bibnamefont {Marcos}},
  \bibinfo {author} {\bibfnamefont {P.}~\bibnamefont {Zoller}}, \ and\ \bibinfo
  {author} {\bibfnamefont {G.}~\bibnamefont {Kirchmair}},\ }\href {\doibase
  10.1103/PhysRevB.92.174507} {\bibfield  {journal} {\bibinfo  {journal} {Phys.
  Rev. B}\ }\textbf {\bibinfo {volume} {92}},\ \bibinfo {pages} {174507}
  (\bibinfo {year} {2015})}\BibitemShut {NoStop}%
\bibitem [{\citenamefont {Fisher}(1964)}]{fisher1964}%
  \BibitemOpen
  \bibfield  {author} {\bibinfo {author} {\bibfnamefont {M.~E.}\ \bibnamefont
  {Fisher}},\ }\href {\doibase 10.1119/1.1970340} {\bibfield  {journal}
  {\bibinfo  {journal} {American Journal of Physics}\ }\textbf {\bibinfo
  {volume} {32}},\ \bibinfo {pages} {343} (\bibinfo {year} {1964})}\BibitemShut
  {NoStop}%
\bibitem [{\citenamefont {Mila}(2000)}]{Mila_2000}%
  \BibitemOpen
  \bibfield  {author} {\bibinfo {author} {\bibfnamefont {F.}~\bibnamefont
  {Mila}},\ }\href {\doibase 10.1088/0143-0807/21/6/302} {\bibfield  {journal}
  {\bibinfo  {journal} {European Journal of Physics}\ }\textbf {\bibinfo
  {volume} {21}},\ \bibinfo {pages} {499} (\bibinfo {year} {2000})}\BibitemShut
  {NoStop}%
\bibitem [{\citenamefont {Giamarchi}(2004)}]{giamarchi2004}%
  \BibitemOpen
  \bibfield  {author} {\bibinfo {author} {\bibfnamefont {T.}~\bibnamefont
  {Giamarchi}},\ }\href {\doibase 10.1093/acprof:oso/9780198525004.001.0001}
  {\emph {\bibinfo {title} {{Quantum physics in one dimension}}}},\
  International series of monographs on physics\ (\bibinfo  {publisher}
  {Clarendon Press},\ \bibinfo {address} {Oxford},\ \bibinfo {year}
  {2004})\BibitemShut {NoStop}%
\bibitem [{\citenamefont {Franchini}(2017)}]{Franchini2017}%
  \BibitemOpen
  \bibfield  {author} {\bibinfo {author} {\bibfnamefont {F.}~\bibnamefont
  {Franchini}},\ }\href {\doibase 10.1007/978-3-319-48487-7} {\emph {\bibinfo
  {title} {{An Introduction to Integrable Techniques for One-Dimensional
  Quantum Systems}}}},\ Lecture Notes in Physics\ (\bibinfo  {publisher}
  {Springer Cham},\ \bibinfo {address} {Switzerland},\ \bibinfo {year}
  {2017})\BibitemShut {NoStop}%
\bibitem [{\citenamefont {Pushpan}\ \emph
  {et~al.}(2024{\natexlab{b}})\citenamefont {Pushpan}, \citenamefont {K~J},
  \citenamefont {Narayan},\ and\ \citenamefont {Pal}}]{Pushpan2024}%
  \BibitemOpen
  \bibfield  {author} {\bibinfo {author} {\bibfnamefont {C.~B.}\ \bibnamefont
  {Pushpan}}, \bibinfo {author} {\bibfnamefont {H.}~\bibnamefont {K~J}},
  \bibinfo {author} {\bibfnamefont {P.}~\bibnamefont {Narayan}}, \ and\
  \bibinfo {author} {\bibfnamefont {A.~K.}\ \bibnamefont {Pal}},\ }\href
  {\doibase 10.1103/PhysRevA.110.032408} {\bibfield  {journal} {\bibinfo
  {journal} {Phys. Rev. A}\ }\textbf {\bibinfo {volume} {110}},\ \bibinfo
  {pages} {032408} (\bibinfo {year} {2024}{\natexlab{b}})}\BibitemShut
  {NoStop}%
\bibitem [{\citenamefont {Barouch}\ \emph {et~al.}(1970)\citenamefont
  {Barouch}, \citenamefont {McCoy},\ and\ \citenamefont
  {Dresden}}]{Barouch1970}%
  \BibitemOpen
  \bibfield  {author} {\bibinfo {author} {\bibfnamefont {E.}~\bibnamefont
  {Barouch}}, \bibinfo {author} {\bibfnamefont {B.~M.}\ \bibnamefont {McCoy}},
  \ and\ \bibinfo {author} {\bibfnamefont {M.}~\bibnamefont {Dresden}},\ }\href
  {\doibase 10.1103/PhysRevA.2.1075} {\bibfield  {journal} {\bibinfo  {journal}
  {Phys. Rev. A}\ }\textbf {\bibinfo {volume} {2}},\ \bibinfo {pages} {1075}
  (\bibinfo {year} {1970})}\BibitemShut {NoStop}%
\bibitem [{\citenamefont {Barouch}\ and\ \citenamefont
  {McCoy}(1971{\natexlab{a}})}]{Barouch1971}%
  \BibitemOpen
  \bibfield  {author} {\bibinfo {author} {\bibfnamefont {E.}~\bibnamefont
  {Barouch}}\ and\ \bibinfo {author} {\bibfnamefont {B.~M.}\ \bibnamefont
  {McCoy}},\ }\href {\doibase 10.1103/PhysRevA.3.786} {\bibfield  {journal}
  {\bibinfo  {journal} {Phys. Rev. A}\ }\textbf {\bibinfo {volume} {3}},\
  \bibinfo {pages} {786} (\bibinfo {year} {1971}{\natexlab{a}})}\BibitemShut
  {NoStop}%
\bibitem [{\citenamefont {Barouch}\ and\ \citenamefont
  {McCoy}(1971{\natexlab{b}})}]{Barouch1971a}%
  \BibitemOpen
  \bibfield  {author} {\bibinfo {author} {\bibfnamefont {E.}~\bibnamefont
  {Barouch}}\ and\ \bibinfo {author} {\bibfnamefont {B.~M.}\ \bibnamefont
  {McCoy}},\ }\href {\doibase 10.1103/PhysRevA.3.2137} {\bibfield  {journal}
  {\bibinfo  {journal} {Phys. Rev. A}\ }\textbf {\bibinfo {volume} {3}},\
  \bibinfo {pages} {2137} (\bibinfo {year} {1971}{\natexlab{b}})}\BibitemShut
  {NoStop}%
\bibitem [{\citenamefont {Sachdev}(2011)}]{Sachdev2011}%
  \BibitemOpen
  \bibfield  {author} {\bibinfo {author} {\bibfnamefont {S.}~\bibnamefont
  {Sachdev}},\ }\href@noop {} {\emph {\bibinfo {title} {Quantum phase
  transitions}}}\ (\bibinfo  {publisher} {Cambridge University Press,
  Cambridge},\ \bibinfo {year} {2011})\BibitemShut {NoStop}%
\end{thebibliography}%

\end{document}